\documentclass[prd,showpacs,superscriptaddress]{revtex4}
\usepackage{epsfig}

\begin{document}     

\title{Flavor  Composition  and Energy Spectrum 
of Astrophysical Neutrinos}

\author{Paolo Lipari}
\email{paolo.lipari@roma1.infn.it}
\affiliation{INFN and Dipartimento di Fisica, 
Universit\`a di Roma I, 
P.A.Moro 2, 00185 Roma, Italy}

\author{Maurizio Lusignoli}
\email{maurizio.lusignoli@roma1.infn.it}
\affiliation{INFN and Dipartimento di Fisica, 
Universit\`a di Roma I, 
P.A.Moro 2, 00185 Roma, Italy}

\author{Davide Meloni}
\email{davide.meloni@roma1.infn.it}
\affiliation{INFN and Dipartimento di Fisica, 
Universit\`a di Roma I, 
P.A.Moro 2, 00185 Roma, Italy}


\begin{abstract}
The   measurement of the flavor composition of 
the neutrino  fluxes
from astrophysical  sources has  been proposed  as a
method to study    not only the 
nature of their emission  mechanisms, but also
the  neutrino fundamental properties.
It is however problematic  to reconcile these  two
goals, since a  sufficiently accurate
understanding of the neutrino  fluxes at the source
is  needed  to  extract information  
about the physics of   neutrino  propagation.
In this  work we  discuss 
critically the expectations for 
the flavor composition   and energy spectrum 
from different  types of astrophysical  sources, and comment on 
the  theoretical  uncertainties
connected to  our limited knowledge of their  structure.
\end{abstract}



\maketitle

\section{Introduction}
There is the  expectation that in the  near future
we will  see the opening  of the  new   field of
observational   high energy neutrino astronomy
\cite{nuh,nuh1,lip-catania}.
Theoretically there are very robust  reasons  to expect
the existence of high energy ($E_\nu \gtrsim 10^{12}$~eV) 
neutrino  sources.  The strongest motivation is the  observation
of  a cosmic  ray  flux,
mostly composed of protons  and  fully ionized  nuclei
 that extends  in energy up to  $E \sim 10^{20}$~eV.
These hadronic  particles  can interact inside or near their
acceleration sites or  during their propagation
in interstellar or intergalactic space.  These interactions
produce  weakly decaying  particles
(such as  $\pi^\pm$  and kaons)  that  generate  neutrinos.
These  ``astrophysical  neutrinos''  are intimately connected
with the high energy photons that  are  created  in the decay
of $\pi^\circ$ and $\eta$ particles  produced in the same  
hadronic  primaries interactions, or in the radiation  processes
of  relativistic  electrons and positrons  co--accelerated 
in  the sources.  A  rich  variety of  high energy  gamma  ray sources
has been observed  with detectors on 
satellites  \cite{egret-3rd} and  ground--based 
\cite{hess-scan} Cherenkov telescopes,
suggesting   several  possible   neutrino sources.

Observations with  neutrinos 
have the  potential  to give us  unique   information about
their astrophysical sources, and hopefully 
could also  result in the discovery of 
new  classes of  sources.
On the other  hand  the possibility to use these observations
to obtain  information about the fundamental properties of the 
neutrinos has been  widely discussed
 \cite{nuh,double-bang,Athar:2000yw,Ahluwalia:2001xc,Pakvasa:2004hu,Barenboim:2003jm,Beacom-flavor}.
Astrophysical neutrinos   travel  pathlengths 
of order $10$~Kpc  ($\sim 3 \times 10^{22}$~cm) for galactic sources,
and as large as  $\sim 1$~Gpc  ($\sim 3 \times 10^{27}$~cm)   
for extragalactic sources.
These remarkably long 
baselines  allow  the  study  of 
phenomena  such as  $\nu$ flavor transitions
or $\nu$ decay  in  a range of parameters  that is  unaccessible
with other methods. Only  observations with SuperNova neutrinos,
very likely   only observable from 
galactic  sources,  but having  much smaller  energy
($E_\nu \sim 10$~MeV)  could 
provide  larger $L/E_\nu$.

In order to extract  
information about the neutrino  fundamental properties
from the observations one needs to have
sufficiently  good understanding of the properties of the 
neutrino emission at the source.
The common assumption  in  several studies of this type
is that the properties of the $\nu$ emission, and in particular
the flavor composition at the  source can   be robustly 
predicted.  
There is  of course a well known  historical  precedent  for  the 
successful use of this  concept
in the   discovery of neutrino oscillations
with  atmospheric  neutrinos.
In fact, the  first  hints of  the existence of
the now  solidly established  
 $\nu_\mu \leftrightarrow \nu_\tau$  transitions  
were obtained   with    the Kamiokande \cite{Kamiokande} 
and IMB  \cite{IMB} detectors
as  the measurement  of a    $\mu/e$  ratio  
 for contained  events  
smaller than the  expectations.
It is natural to  try to make use 
of  new, very distant neutrino   sources
(when they  will be  discovered) 
to perform  additional   studies  of   the properties
of neutrino propagation.  

Several   ``exotic''  processes, beyond
standard flavor oscillations,   could 
reveal  themselves only 
in the propagation of   neutrinos
over  very long distances.
For example, it has  been suggested
that some neutrinos could decay into  a lighter neutrino  and a
majoron  \cite{majoron}, if the lifetime 
is sufficiently  long 
this phenomenon could  be only detectable 
for neutrinos propagating over astronomical  distances
\cite{Acker:1991ej,nu-decay}.
A  second interesting  possibility 
is  that neutrinos are
pseudo--Dirac states \cite{pseudoDirac}
where  each generation is actually composed of two 
maximally-mixed Majorana neutrinos separated by a tiny mass difference.
If the  pseudo-Dirac splittings   are  sufficiently small,
the phenomenology of  oscillations on  short baselines
remains  unchanged, however  
when $E_\nu/L$  
becomes  comparable or smaller  than the 
 pseudo--Dirac splittings   new  transitions  become possible,
and can in principle  be  detectable with astrophysical neutrinos
\cite{beacom-pd}.  More in general,  oscillations
into  sterile states that are quasi  degenerate
to the active neutrinos  can in principle be investigated 
down  to very small  squared mass  splittings \cite{steriles}.
Several  other  mechanisms    such as
 quantum decoherence \cite{Hooper:2004xr}, 
violations of the equivalence principle 
\cite{Gasperini:1989rt,Minakata:1996nd}, neutrinos  with varying masses
\cite{Fardon:2003eh,Hung:2003jb}  could leave  their signature
on the propagation of astrophysical neutrinos.

A well known ``naive'' argument states  that
since  the dominant source of  astrophysical neutrinos
is the decay of  charged pions,  and each   $\pi^\pm$,
after  chain decays  such as:
$\pi^+ \to \mu^+ \nu_\mu 
\to (\overline{\nu}_\mu \, e^+ \,  \nu_e) \, \nu_\mu$, 
generates   two muon neutrinos
and one   electron neutrino,
 the  flavor  ratio at the source is 
$R_{\mu e} = (\nu_\mu + \overline{\nu}_\mu)/(\nu_e  + \overline{\nu}_e) \simeq 2$.
This  naive argument  has some pedagogical value, 
but must  be considered only  as a  first approximation.
In fact, even  if  charged pions are the only source
of neutrinos,    $R_{\mu e}$ is only  equal to 2
after integration over all  $\nu$  energies,
because  the three neutrinos  generated in a  charged pion decay
have    different  energy spectra,
and  therefore in general the   flavor ratio 
will depend on the spectral shape 
 of the  $\nu$   signal, and  will  vary with the 
neutrino  energy.
An effect that can be of large   importance
is the presence of  sufficiently efficient
energy loss  mechanisms in the source. 
In this situation the 
particles that are  most affected
are the muons.  The presence of 
additional neutrino sources 
(such as kaons)  can also be
important for
the flavor  ratio.
In this work   we will re--examine  critically   the
uncertainties in  the  predictions of the spectra 
and flavor  composition
of astrophysical neutrinos  and discuss the  implications
for the  extraction of information on the neutrino properties.

This paper is  organized as follows:
in the next section 
we  discuss  how  the observable  neutrino flavor ratios
carry information  at the same time 
about the  flavor  composition at the source
and about  the flavor  transition probabilities.
In  section~3 we outline 
the general structure of the  calculation of the 
neutrino  signal  from an astrophysical  source.
The  different steps  of these calculations
are considered in more  details in sections 4--7.
In section~8, as an  illustration,
we describe
an explicit calculation of the   neutrino  emission
from the fireballs  of Gamma Ray Bursts 
following the model  of  Waxman and Bahcall \cite{Waxman-grb}.
Section~9  contains a  short  discussion of the problem
of  the experimental determination 
of the  flavor  composition and energy spectra 
of a neutrino signal. The last section  presents our conclusions.

\section{Flavor  Ratios  of Astrophysical Neutrinos}
The observable  fluxes of astrophysical neutrinos   
from a source at distance $L$ will be
linear combinations of  the fluxes  at the source.
Leaving  implicit the  energy and distance  dependences
one  can write:
\begin{equation}
\phi_{\nu_\alpha} = \sum_{\nu_\beta} 
P_{\nu_\beta \to \nu_\alpha} ~\phi_{\nu_\beta}^\circ
~~.
\end{equation}
The  transition probabilities $P_{\nu_\beta \to \nu_\alpha}$
have certainly a  non trivial  structure  because  of the
known existence   of  ``standard'' flavor oscillations,
 but  might   depend on
additional ``new physics'' contributions,  such
as neutrino decay, that  become  significant only for
very long pathlengths.
In the standard  scenario  the neutrino   number is
conserved  and therefore $\sum_\beta P_{\alpha \to \beta}  = 1$.
More in general, in the presence
of a  non negligible decay probability or of
transitions to additional sterile states, the  sum  
can be  less than unity.

The  flavor, energy and pathlength  dependences
of the  standard oscillation
probabilities are   well known,  
and are   determined by  
two squared  mass  differences
($\Delta m^2_{12} \simeq 8.0 \times 10^{-5}$~eV$^2$, 
$|\Delta m^2_{23}| \simeq 2.5 \times 10^{-3}$~eV$^2$)
and the neutrino  mixing  matrix.
The  oscillation lengths 
($\lambda_{ij} = 4\pi \,E_\nu/|\Delta m^2_{ij}|$)
of the standard flavor  transitions
are short    with respect to 
the typical  astrophysical  distances
and, in most cases, it is  a good approximation
to consider only the probability   averaged over
distance. In this case the probability  becomes 
independent from  $E_\nu$ and $L$  and  takes the form:
\begin{equation}
\langle P_{\nu_\alpha \to \nu_\beta}^{\rm standard} \rangle_{L} =
\sum_j | U_{\alpha j}|^2 \,  | U_{\beta j}|^2
\end{equation}
where $U$ is the unitary mixing matrix  that 
relates the neutrino   flavor
and  mass eigenstates. This  matrix   can be written 
in terms of three mixing  angles and  one  CP violating phase.
From  a global fit   to all existing  neutrino data 
\cite{Strumia:2005tc}
one  can extract
the best fit   values:
$\theta_{12} \simeq 34^\circ$, 
$\theta_{23} \simeq 45^\circ$, 
$\theta_{13} \simeq 0$,
and  99\% C.L. intervals for the mixing angles:
$\theta_{12} \in [30^\circ, 38^\circ]$, 
$\theta_{23} \in [36^\circ, 54^\circ]$, 
$\theta_{13} \le 10^\circ$;
the phase $\delta$ remains  completely undetermined.

Assuming  to know  sufficiently  well the   properties
of the source,  the  observed  flavor ratios  can  give
information   on the  flavor transition probabilities.
This can  be used to help in the determination of the ``standard''
parameters in the  flavor  oscillations, 
or more ambitiously to    investigate the possible existence 
of additional  phenomena   in neutrino propagation.

\begin{figure} [hbt]
\centerline{\psfig{figure=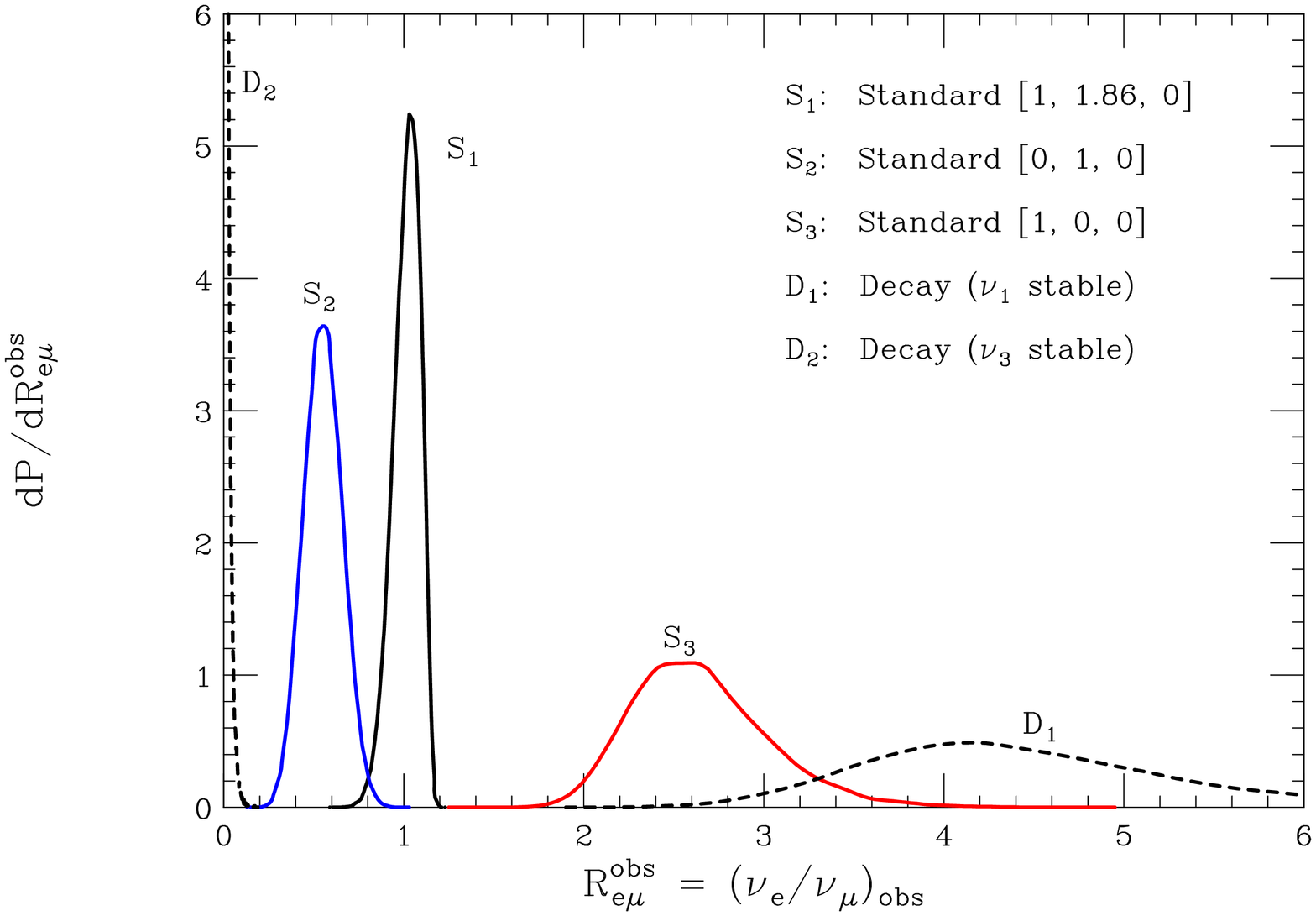,angle=90,width=11.cm}}
\caption {
Distribution of  the observable flavor  ratio 
$R_{e\mu} = (\nu_e/\nu_\mu)_{\rm obs}$
for    different  models  of the $\nu$ source and
different neutrino  properties. 
A  source  model 
is defined by the relative  intensity  
(summing over $\nu$ and $\overline{\nu}$) 
of the emission for the three 
flavors: [$\nu_e,\nu_\mu,\nu_\tau]_{\rm source}$.
The distribution  of $R_{e \mu}$ is  entirely
due to the present  uncertainties
in  the determination of the    $\nu$ mixing parameters.
The  source  models 
$[0,1,0]$, and $[1,0,0]$ correspond  to 
pure $\nu_\mu$ or $\nu_e$ emission;
the model  $[1,1.86,0]$ corresponds to
the pion dominated emission
from a  thin source  with a power  law  spectrum of slope 2.
The two  dashed   curves  
show the  distributions of $R_{e \mu}$ 
for  a  $\nu$--decay model  
\protect\cite{nu-decay}
 where  only the  lowest $\nu$  mass   eigenstate 
(in the direct and inverse hierarchy)
is  stable.
\label{fig:flav1}  }
\end{figure}

\begin{figure} [hbt]
\centerline{\psfig{figure=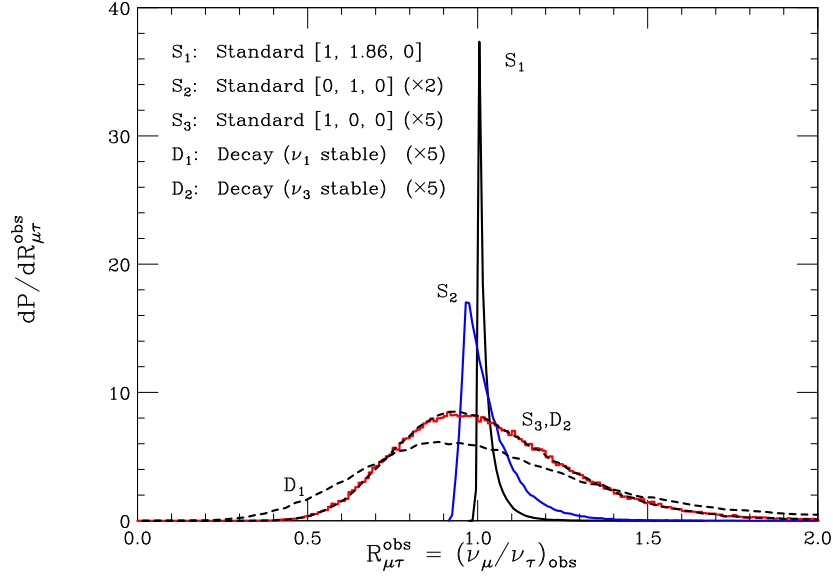,angle=90,width=11.cm}}
\caption {
Distribution
of the observable flavor  ratio 
$R_{\mu\tau}= (\nu_\mu/\nu_\tau)_{\rm obs}$
(see caption of fig.~\ref{fig:flav1}).
Some  of the curves have been renormalized
for an easier  observation of their shape.
\label{fig:flav2}  }
\end{figure}

As an illustration in  fig.~\ref{fig:flav1}  and \ref{fig:flav2}
we show the expectations for the  two  independent flavor  ratios  
$R_{e\mu} = (\nu_e/\nu_\mu)_{\rm obs}$
and $R_{\mu\tau} = (\nu_\mu/\nu_\tau)_{\rm obs}$,
calculated    for  different  assumptions
for  the source emission, in the presence
of simple standard oscillations,  or also including  
neutrino  decay  (with  two different assumptions)
   \cite{nu-decay}.
The distributions
of the flavor ratio for a given model  are determined  only by the
present  uncertainties   on the  neutrino
mixing  parameters.
A  source model  is   defined 
by the relative  intensity  of the emission for the three
neutrino  flavors.
For our  illustration we have considered  three source models.
The first model is the  emission of
form: [$\nu_e,\nu_\mu,\nu_\tau]_{\rm source} = [1,1.86,0]$.
This  corresponds  (as we will discuss more extensively in the
following) to  pion dominated emission
from a    thin source (no  significant 
energy loss for secondaries)
with a power  law  energy spectrum of slope 2.
Several classes  of  $\nu$ sources of this  type
(such as  Supernova Remnants and Gamma Ray  Bursts)
have been predicted.
Note that  the  flavor  relative   abundances in this model 
are  close  but not identical to
the  naive  expectation  [1,2,0]. 
The deviations 
are  a  simple  consequence of the shape  of the
energy spectra of the neutrinos  in  pion  chain decay.
The   observable $R_{e\mu}$ ratio   for this  model
takes value  in the interval $\simeq (0.7,1.2)$
with a most likely value close to unity, while
the other independent ratio $R_{\mu\tau}$  has a narrow
distribution sharply peaked at the value 1, 
with a tail extending    up to $\simeq 1.3$. 

Figures~\ref{fig:flav1}  and~\ref{fig:flav2}
also show the predicted   flavor  ratios for  emission
with (summing  over $\nu$ and $\overline{\nu}$) flavor  abundances
$[0,1,0]$, and $[1,0,0]$, that is 
pure  muon  or electron neutrino emission.
The motivation   for using  these two  models 
is that, to a very good approximation,
they can be considered the  two extreme   models 
for the  emission from  a ``standard'' (not  involving new physics) 
source, since  a  significant production of
$\nu_\tau$  in an astrophysical   environment 
is  extraordinary  unlikely.
Sources   emitting pure
fluxes of  $\nu_\mu$ or  $\nu_e$  are in principle possible,
and  have been in  fact advocated in the  literature.
They correspond  to a
source dominated by pions   where  
muons  lose  all their  energy before decay,
or to a pure neutron  source, that  generates
a $\overline{\nu}_e$  flux.  
A pure  neutron  ($\overline{\nu}_e$)  neutrino  source is
in  fact unrealistic  since  it is  natural that
neutron  production is associated with
 pion production and therefore  with some $\nu_\mu$
emission.  
We  will comment more extensively on 
these issues in  the  following.

For the [0,1,0]  ([1,0,0]) model the
most likely value    for the observable flavor  ratio is 
 $R_{e\mu} \simeq 0.55$ 
($R_{e\mu} \simeq 2.7$).
In both cases  
 $R_{\mu\tau}$  is  centered at  the value  unity,
this  is a simple  consequence of the  large (approximately maximal)
mixing between   $\nu_\mu$ and  $\nu_\tau$.
In the case of  a [1,0,0]  source  the $R_{\mu\tau}$ distribution is
noticeably broader  extending  
to the interval $R_{\mu \tau} \in (0.5,1.5)$.

Neutrino  decay  with a very long lifetime is one of the
most interesting possibilities that can  be investigated with
astrophysical  neutrinos. 
Assuming    \cite{nu-decay,Barenboim:2003jm}
   that only  the 
lowest mass  eigenstate is  stable, 
and that the  distance of the source
is  much longer  than the decay lengths,
the    observable flavor  ratios 
depend  uniquely on the  flavor composition of 
the  lightest  eigenstate.  Depending
on the  sign  of $\Delta m^2_{23}$
this is  
 $\nu_1$ (direct  mass hierarchy)
or  $\nu_3$ (inverse  hierarchy).
The  flavor  ratios  for these two models 
are also shown in fig.~\ref{fig:flav1} and~\ref{fig:flav2}
as dashed lines. The $R_{e \mu}$  ratio  is  particularly
interesting. In the case of  direct mass  hierarchy,
the most probable value 
of the   $e\mu$ ratio   is large
($R_{e\mu}\simeq 4.2$),
the  distribution is  also remarkably wide, allowing in principle
to   constrain  the  mixing  parameters \cite{nu-decay}.
For the inverse hierarchy, when 
the  stable  neutrino  is the $\nu_3$  that has a small
(or  perhaps vanishing)  overlap with  $\nu_e$,
the observable $R_{e \mu}$ is  close to zero.

It is    remarkable    that   the predictions
for the $e\mu$ flavor  ratio in   neutrino decay
are in both (direct and inverse  mass hierarchy) 
models  significantly different   from  the standard   model
predictions   for    nearly   all    assumptions
about the  flavor  composition of the emission.
Only a pure $\nu_e$ emission has some  overlap with the 
decay  model with stable $\nu_1$.
The  study of  the $e\mu$ flavor ratio
of astrophysical  neutrinos  can therefore
give evidence in favor or against the  existence 
of  $\nu$ decay.

The determination of the  neutrino mixing parameters
in the   standard oscillation  scenario is more problematic.
For   the most sensitive $e\mu$  ratio, 
the range of possible  values 
due to the present uncertainties on the mixing parameters
is  in  fact smaller  than the  
variations   that   result from different  flavor
abundances at the source. Therefore  the possibility
to obtain   interesting bounds on the mixing parameters
depends crucially  on  having a sufficiently precise
knowledge of the source.  

To  illustrate this  problem, 
we can  write  an approximate expression for 
the flavor ratio  $R_{e \mu}$ in      terms of
the relevant mixing matrix  parameters
expanding in first order around the best fit values
\begin{equation}
\left (
\frac{\nu_e}{\nu_\mu} 
\right )_{\rm obs}^{[1,1.86,0]}
 \simeq    1.026  
- 0.0218 ~ \delta \theta_{23}^\circ
+ 0.0087 ~ \theta_{13}^\circ \; \cos \delta
- 0.0014 ~ \delta \theta_{12}^\circ
+ 0.370 ~\delta \left (
\frac{\nu_e}{\nu_\mu} 
\right )_{0} ~~~
\label{eq:rflav1}
\end{equation}
\begin{equation}
\left (
\frac{\nu_e}{\nu_\mu} 
\right )_{\rm obs}^{[0,1,0]}
 \simeq   0.547  
- 0.0296 ~ \delta \theta_{23}^\circ
+ 0.0119 ~ \theta_{13}^\circ \; \cos \delta
+ 0.0197 ~ \delta \theta_{12}^\circ
+ 1.152 ~\delta \left (
\frac{\nu_e}{\nu_\mu} 
\right )_{0}~~~
\label{eq:rflav2}
\end{equation}
\begin{equation}
\left (
\frac{\nu_e}{\nu_\mu} 
\right )_{\rm obs}^{[1,0,0]}
 \simeq   2.65  
+ 0.093 ~ \delta \theta_{23}^\circ
- 0.037 ~ \theta_{13}^\circ \; \cos \delta
- 0.131 ~ \delta \theta_{12}^\circ
- 3.846 ~\delta \left (
\frac{\nu_\mu}{\nu_e} 
\right )_{0}~~~
\label{eq:rflav3}
\end{equation}
In  these equations 
$\delta\theta_{12}^\circ$ 
and $\delta\theta_{23}^\circ$ are   the 
deviations  in degrees 
 of the mixing  angles from their best fit value
($\theta_{12} = 34^\circ$,
$\theta_{23}  = 45^\circ$),
$\theta_{13}^\circ$  is the value (always in degrees)
of the third  angle
and $\delta$ is the CP  violating phase.
The superscript labels  
of the flavor  ratio  indicate  the source  model.
The last term   in  equations
(\ref{eq:rflav1}),
(\ref{eq:rflav2}) and
(\ref{eq:rflav3})   gives
an estimate of
the   variation in the observable   ratio  due
to  the  uncertainty in the flavor ratio
at the source obtained
expanding   in  first order in 
$\delta(\nu_\alpha/\nu_\beta)_0$ around   the flavor
abundances of the model considered.
One  can see that    uncertainties  in  
flavor  abundances at
the source  can  be so large to make impossible a
meaningful  constraint on the  mixing parameters.

In the remaining  sections of this paper we will discuss
how the properties of the astrophysical  sources
determine the flavor abundances of their emission,  and  
estimate the  associated theoretical  uncertainties.

\section{Production of Astrophysical Neutrinos}
Astrophysical  neutrinos are generated 
when a population of  relativistic hadrons  
(protons  or nuclei)  interacts with 
some  target  material  (gas or a radiation field)
inside, near,  or outside the  acceleration site. 
These interactions  produce weakly decaying 
secondary particles    
whose decay generate neutrinos  either directly, or indirectly
with the subsequent decay of  muons.
The  description of these processes  therefore requires the following
elements:
\begin{enumerate}
\item  The description of the    energy spectrum 
and composition of the primary particles.

\item  The   definition  of the   target  material 
with  which the primary particles  interact.

\item  The modeling  of the properties of 
particle production in hadronic  interactions.

\item  The  description  of the   properties of the medium where
the interactions are  taking place, to determine  the 
relevant mechanisms  for  energy loss.

\item  The (well known)  properties of weak decays.

\end{enumerate}
In our  discussion  we will   describe the source
as one   stationary homogeneous  region.
This  is an  important limitation, because in 
general  we can expect that the 
sources  will  have   non trivial 
space structures,
 with  the neutrinos   emerging   from different  regions that  can 
produce different spectral shapes   and  flavor contents;
most high  energy radiation sources are also expected    to  show
important time  variabilities.
Because of the limited  statistics  and  angular resolution,
the  observations of astrophysical  neutrinos
will necessarily  integrate  over the entire volume
of the source (that in most cases will appear as point--like)
and average over most or all  time  variations.
Describing this situation requires 
the averaging   over 
different  source  regions and conditions.

Some of  the most interesting  postulated sources
of  neutrinos, in particular  
the jets of Active Galactic  Nuclei (AGN)
and   Gamma Ray Bursts (GRB),  are  expected to be
associated   with  astrophysical  jets 
with   ultra--relativistic   bulk motion. 
These  relativistically moving sources  
are best  described  in the rest frame of
the jet, where the   emitted  radiation is 
approximately  isotropic, and the  electromagnetic fields
can  be described as a purely magnetic  field.
The observable   fluxes can then be obtained
by an appropriate Lorentz boost.

The  $\nu$ emission from  a  given  source
can be  described by the  emissivities
 $Q_{\nu_\alpha} (E_\nu, \Omega)$   (in units
(s~sr~GeV)$^{-1}$)
that    give the number
of neutrinos   of type $\alpha$  emitted   
with energy $E_\nu$ in the  direction $\Omega$. 
The  $\nu$ fluxes at  the Earth are   obtained
from the   neutrino  emissivities as:
\begin{equation}
\phi_{\nu_\alpha} (E_\nu) = 
\frac{(1+z)^2}{4 \pi d_L^2} ~
\sum_{\nu_{\beta}} P_{\nu_\beta \to \nu_\alpha}(E_\nu) 
~Q_{\nu_\beta}  [E_\nu (1+z),\Omega_\oplus]
\label{eq:q-flux}
\end{equation}
where $d_L$ is the source  luminosity distance,
$z$ is   its redshift, and   $ P_{\nu_\beta\to \nu_\alpha}$
are  the flavor  transition probabilities.
The observable  fluxes depend  only on the
emission in   the direction $\Omega_\oplus$ 
that corresponds to the line of sight from the source to the Earth.  
In the following we will not  consider
the  angular dependence of the source emission,
since this is  not  observable   for any individual source.

The  calculation of the neutrino emissivities $Q_{\nu_\alpha} (E_\nu)$,
starts  from a decription of the
energy spectrum  and composition of the primary particles. 
In the following we will  assume that 
the  primary particles are protons  and are   described
by the  energy spectrum  $N_p (E_p)$  (in units GeV$^{-1}$).
The  extension to the case  where  there is a significant
contribution from   nuclei 
is straightforward, and to  a  very good approximation
can be   calculated  introducing   effective  proton
and neutron   spectra ($N_p (E_p)$ and $N_n (E_n)$) 
that  take also into   account the  bound  nucleons. 
The  description of the  environment  of the source  
must  include the density and properties of the target material
and of all other  fields   that   can be a source
of  energy loss.
The interactions of the  primary particles
result in the emission of    secondary  particles
$(\pi^+, \pi^-, K^+, \ldots$)  
that are sources of  neutrinos.  These particles
must be   propagated in the source  medium  until
they decay.  All muons  generated by these decays must 
also  be propagated and   their decay studied.

The scheme  of the   calculation is  therefore the following.
The rate of production of secondary particle 
of  type $a$  can be calculated   with the convolution:
\begin{equation}
Q_a (E_a) = \int  dE_p  ~N_p (E_p) ~ K_{p} (E_p) ~~
\frac{dn_{p \to a}}{dE_a}  (E_a; E_p)
\end{equation}
where  $K_{p} (E_p)$ is the  interaction probability per unit time
of a proton  of energy $E_p$, and 
$dn_{p \to a}/dE_a (E_a; E_p)$ is the  
number of particles of type $a$ in the final state
in the energy interval ($E_a$, $E_a + dE_a$).

The  energy distribution of particle $a$ at  decay 
$Q_a^{\rm dec} (E)$ will in  general
differ from the distribution at production and can be obtained
with the integration:
\begin{equation}
Q_a^{\rm dec} (E) = \int_{E}^\infty  dE_i~Q_a (E_i)~
\frac{dp_{\rm dec}^a}{dE}  (E; E_i) .
\label{eq:edec}
\end{equation}
where $dp_{\rm dec}^a/dE (E; E_i)$ is the probability  density
that particle  $a$   created   with energy $E_i$ will decay
with energy $E$.  

The neutrinos  created in the {\em direct}  decay of particle $a$ can
be  calculated as:
\begin{equation}
Q_{a \to \nu_\alpha} (E_\nu) = \int_{E_\nu}^\infty dE_a ~
Q_a^{\rm dec} (E_a) ~ \frac{dn_{a \to \nu_\alpha}} {dE_\nu} (E_\nu;
E_a) 
\label{eq:adec}
\end{equation}
folding the parent energy distribution
(at  decay)  with  the appropriate decay spectra
$dn_{a\to \nu}/dE_\nu (E_\nu; E_a)$.

The   muon production rate  can be calculated
with an expression   that is  the analogous of
(\ref{eq:adec})  substituting the appropriate decay
distributions:
\begin{equation}
Q_{\mu} (E_\mu) = \sum_a ~
\int_{E_\mu}^\infty dE_a ~
Q_a^{\rm dec} (E_a) 
~ \frac{dn_{a \to \mu}} {dE_\mu} (E_\mu;
E_a) 
\label{eq:qmu}.
\end{equation}
The sum runs over all weakly decaying  hadrons.

In out calculation we  explicitely included  charged  pions, kaons  and
neutrons,  neglecting the much  smaller contributions
of heavier particles.
An important   complication is that the 
spectra of  the particles   generated in muon decay
depend on the muon  helicity
 (see discussion in section~\ref{sec:weak}).
It is therefore  convenient to
calculate  separately the production of muons 
 with different  helicity:
\{$\mu^+_L$, 
$\mu^+_R$, 
$\mu^-_L$ and  
$\mu^-_R$\}  (where $L$ and $R$ indicate the helicity).
It is expected  that the  helicity of the muons
is  to a  good approximation conserved  even in the presence
of a strong magnetic  field and of  significant energy losses, and
therefore the    energy distributions 
of the muon   of different helicities  at decay
can be  obtained   with a convolution similar to  (\ref{eq:edec}):
\begin{equation}
Q_{\mu^\pm_h}^{\rm dec} (E) = \int_{E}^\infty  dE_i
~Q_{\mu^\pm_h} (E_i)~
\frac{dp_{\rm dec}^\mu}{dE}  (E; E_i) .
\label{eq:mudec}
\end{equation}
The last  step  is to compute the 
neutrinos  generated     by muon decay:
\begin{equation}
Q_{\mu^\pm_h  \to \nu_\alpha} (E_\nu) = 
\int_{E_\nu}^\infty dE_\mu ~
Q_{\mu^\pm_h}^{\rm dec} (E_\mu) 
~ \frac{dn_{\mu^\pm_h \to \nu_\alpha}} {dE_\nu} (E_\nu;
E_\mu) 
\label{eq:mudec1}.
\end{equation}
The neutrino  flux is  obtained  summing over all possible sources.

In the following sections  we will give more details  about the
neutrino  production.  Our  discussion 
starts  from  the neutrinos and
goes  backward along the chain of processes  that lead to their 
production.
The next section contains a 
discussion of the  decay energy distributions;
after discussing the consequences for
exact power law spectra in section~\ref{sec:power},
in section~\ref{sec:pdec}    we will discuss 
possible mechanisms  for  energy loss and  the
calculation of  the  decay energy distribution;
in section~\ref{sec:inter}
we will discuss  the interaction probability $K_{p} (E_p)$ and
the properties of particle production 
in  hadronic  interactions.

\section{Energy Spectra in weak decays }
\label{sec:weak}
For an accurate  prediction of the neutrino  fluxes it is 
necessary to  include a  precise  description of the
energy spectra  produced in  weak  decays.
The energy distribution of particle $b$ 
in the decay  $a \to b + \ldots$  
in the
frames where  the parent particle  $a$ is  ultrarelativistic,
takes  the  scaling form:
\begin{equation}
\frac{dn_{a \to b}}{dE} (E_b; E_a) 
 = \frac{1}{E_a} 
~ F_{a \to b} \left  (
\frac{E_b}{E_a}\right ).
\label{eq:dec_scaling}
\end{equation}
The  functions 
$F_{a \to b} (x)$  for the relevant 
weak decays  are  calculable   using the measured 
 branching   fractions in the possible  final states,
and their  matrix  elements.

For the two body  decay $\pi^+ \to \mu^+ \nu_\mu$,
that accounts for approximately 100\% of charged pion decay,
the  scaling distributions 
are fully determined  by  elementary kinematics:
\begin{equation}
F_{\pi^+ \to \nu_\mu} (x) 
= {1 \over 1 -r_\pi } ~ \Theta (1-r_\pi -x )
\end{equation}
\begin{equation}
F_{\pi^+ \to \mu^+} (x) = {1 \over 1 -r_\pi } ~ \Theta (x-r_\pi)
\label{eq:pimu}
\end{equation}
where $\Theta (x)$ is the step  function  and  
$r_\pi = (m_\mu/m_\pi)^2$.
It is also necessary to take into account
the spin state  of the  final state muons,
that can be described   by the 
helicity  $h = P_R - P_L$,  where  $P_{R,L}$ is  the probability
that the muon has spin parallel (anti--parallel) to its momentum.
The helicity, 
in a  frame where the  parent pion is  ultrarelativistic,
is a function of the
fractional  energy $x$ \cite{Lipari-lep}:
\begin{equation}
h_{\pi^+ \to \mu^+}(x) 
= \frac{2\,r_\pi}{\left( 1 - r_\pi \right) \,x}
-\frac{1 + r_\pi}{1 - r_\pi} .
\label{eq:pol}
\end{equation}
Because of  $CP$ invariance:
\begin{equation}
h_{\pi^- \to \mu^-}(x) = 
-h_{\pi^+ \to \mu^+}(x)
\end{equation}
The helicity  $h_{\pi^+ \to \mu^+}(x)$ takes the 
value $-1$ for $x \to 1$ (forward muon emission)
and $+1$ for $x \to r_\pi$  
(backward muon emission)  reflecting the fact
that  the $\mu^+$   is  created   as  a left--handed
particle in the  pion rest  frame  
in order to compensate the angular monentum of
a left--handed  neutrino.

A useful  method to take  into account the  effect of muon
polarization  in the presence of   energy loss for the muons
is to consider   separately the production
of left--handed  and right--handed  muons, that have 
two well determined  decay spectra.  Combining equations
(\ref{eq:pimu}) and (\ref{eq:pol})  one  obtains the   distributions:
\begin{equation}
F_{\pi^+ \to \mu^+_R} (x) =
F_{\pi^- \to \mu^-_L} (x) =
 \frac{r_\pi (1-x)}{(1-r_\pi)^2 \, x}
~ \Theta (x-r_\pi)
\label{eq:pimu1}
\end{equation}
\begin{equation}
F_{\pi^+ \to \mu^+_L} (x) =
F_{\pi^- \to \mu^+_R} (x) =
 \frac{x - r_\pi}{(1-r_\pi)^2 \, x}
~ \Theta (x-r_\pi)
\label{eq:pimu2}
\end{equation}
Because of CP invariance one has:
$F_{\pi^+ \to \mu^+_{R,L}} (x) = 
F_{\pi^- \to \mu^-_{L,R}} (x)$.

The above   discussion  
is  also valid  for the decay mode  $K^+ \to \mu^+  \nu_\mu$
(and  charge  conjugate), with the  simple   replacement
$m_\pi \to m_K$.

The scaling functions that  describe the decay  of
a muon of helicity $h$   are:
\begin{equation}
F_{\mu^+ \to \overline{\nu}_\mu} (x; h) = 
\left ( \frac{5}{3} - 3\,x^2 + \frac{4\,x^3}{3} \right )
+ h~ \left (
- \frac{1}{3}  + 3\,x^2 - 
  \frac{8\,x^3}{3}
\right )
\label{eq:mudec2}
\end{equation}
\begin{equation}
F_{\mu^+ \to \nu_e} (x; h) = 
\left (
2 - 6\,x^2 + 4\,x^3
 \right )
+ h~ \left (
2 - 12\,x + 18\,x^2 - 8\,x^3
\right )
\label{eq:mudec3}
\end{equation}
Because of  $CP$  invariance the spectra for the charged  
conjugate decays  are  obtained    with the replacement
$h \to -h$.
The spectra  for  unpolarized muons
can be obtained  setting $h = 0$.

If the  muon energy loss   before  decay is  negligible,
it is   straightforward to convolute  the  previous expressions
to obtain the $\nu$ spectra after a chain decay 
($\pi \to \mu  \to \nu$) or
($K \to \mu  \to \nu$).
In the more  general case the energy loss of the muon  
must be taken into account.

The  energy  spectra of  neutrinos and  muons
emitted 
in the three body decays of  kaons
($K^{\pm,\circ}_{e 3}$  and $K^{\pm,\circ}_{\mu 3}$)
can be written
 in terms  of vector form  factors
that  have been   experimentally determined  \cite{PDG}.

Neutrons  are also a source of $\overline{\nu}_e$.  The energy
spectrum of these neutrinos is also  well known.
The $\overline{\nu}_e$  carry a very 
small fraction  of  the parent  neutron energy
and are negligible in most (but not all) cases.

Several   works  on the fluxes  of 
astrophysical neutrinos  make
simplifying assumptions about the   energy distributions
of the neutrinos  produced in weak  decays. 
For  example the calculation of Kashti and Waxman
\cite{Kashti} approximates the 
decay spectra  of  charged pions and muons into
muons and neutrinos as:
\begin{equation}
F_{\pi^+ \to \nu_\mu} (x) = \delta \left [ x - \frac{1}{4} \right ]
\end{equation}
\begin{equation}
F_{\pi^\pm \to \mu^\pm} (x) = \delta \left [ x - \frac{3}{4} \right ]
\end{equation}
\begin{equation}
F_{\mu^+ \to \overline{\nu}_\mu} (x) =
F_{\mu^+ \to \nu_e} (x) =
 \delta \left [ x - \frac{1}{3} \right ]
\end{equation}
These   approximations, 
together with the assumption that  muon energy loss 
before decay is  negligible, 
imply that a
$\nu$ source  dominated by pions   has
a  flavor ratio  ($\nu_\mu + \overline{\nu}_\mu)/(\nu_e +\overline{\nu}_e) = 2$.
With the use of the  correct  expressions 
(\ref{eq:pimu1}), 
(\ref{eq:pimu2}), 
(\ref{eq:mudec2}) and
(\ref{eq:mudec3})  for the decay spectra one obtains
the result that   even for a 
pure pion source  the flavor ratio  
is   not  exactly two, and is 
in  general a function of the $\nu$   energy
 determined  by the  shape  of the neutrino spectrum.

\section{Power law Spectra}
\label{sec:power}
The situation where the   neutrino energy spectrum  
is  a power  law  of form $\phi_\nu \propto  E_\nu^{-\alpha}$ 
is  phenomenologically very important, 
and  is  simple to  discuss.  A power  law  $\nu$ spectrum 
implies that  the  spectrum of the parent particles
(pions and kaons) is also a  power law  with the same  slope
($\alpha_{\pi, K} =\alpha_\nu$).
Such a spectrum 
arises   when  the interacting primary
particles   have  a power law spectrum
and the target is  either composed of normal  matter at rest,
or  is  a photon field  with an energy distribution  that 
has  again a power law  form.
In the first case the slope of the  secondary particles
spectrum is equal to the one for  the primary particles
($\alpha_{\pi,K} \simeq \alpha_p$);
 in the second  case the 
slope of the secondaries is $\alpha_{\pi,K} \simeq
 \alpha_p - \beta +1$
where   $\alpha_p$ the slope of the primary particles and
$\beta$ the slope of the target photon distribution 
($n_\gamma \propto \varepsilon^{-\beta}$).

In general the spectrum of  neutrinos 
produced in the decay  of   particle  type  $a$ 
can be obtained 
convoluting  the  energy spectrum of the
parent  particles  with the  appropriate 
weak decay spectrum
 (see equation (\ref{eq:adec})). 
 If the parent  particle   
spectrum has a power  law  form,  because of the scaling
form  (\ref{eq:dec_scaling})
of the decay spectra  the resulting 
neutrino   energy distribution  can  be written as:
\begin{eqnarray}
Q_\nu (E_\nu) & = & 
\int_{E_\nu}^{\infty} ~dE_a ~Q_a (E_a) ~
{dn_{a \to \nu} \over dE_\nu } (E_\nu ; E_a) \nonumber  \\
& = & 
\int_{E_\nu}^{\infty} ~dE_a ~\left (C_a \; E_a^{-\alpha} \right )~
{1 \over E_a} ~
 F_{a \to \nu}\left  ( {E_\nu \over E_a } \right )
\nonumber \\
& = & 
C_a \;  E_\nu^{-\alpha} ~\int_0^1 ~dx ~x^{\alpha -1} 
~ F_{a \to \nu}(x) 
= C_a \; Z_{a \to \nu}  (\alpha)\; E_\nu^{-\alpha}.
\label{eq:zz}
\end{eqnarray}
In other words the neutrino spectrum is a power
law  with the same  slope
of the parent particles, with  a proportionality   factor
that is commonly called the ``$Z$--factor'', and is 
the momentum of order $\alpha-1$  of the 
decay spectrum: 
\begin{equation}
 Z_{a \to \nu_j} (\alpha) = \int_0^1 dx~x^{\alpha -1} 
~F_{a\to \nu_j}(x).
\end{equation}
$Z_{a \to \nu_j} (1)$  is  the
average   multiplicity  of $\nu_j$
in the final state,
 and $Z_{a \to \nu_j} (2)$   is the
fraction of the parent particle energy 
carried  away  by neutrinos of type $j$.

In the case of $\pi^\pm$ chain 
decay  the  three    relevant $Z$--factors
are:
\begin{equation}
Z_{\pi^+ \to  \nu_\mu} (\alpha) =
\frac{(1- r_\pi)^{\alpha-1}}{\alpha}
\end{equation}
\begin{equation}
Z_{\pi^+ \to \mu^+ \to \overline{\nu}_\mu} (\alpha) =
\frac{4\,\left( 3 - 2\,r_\pi - 
      \left( 3 + \alpha \right) \,r_\pi^\alpha + 
      \left( 2 + \alpha \right) \,r_\pi^{1 + \alpha} \right)
      }{\alpha^2\,\left( 2 + \alpha \right) \,
    \left( 3 + \alpha \right) \,
    {\left( 1 - r_\pi \right) }^2}
\end{equation}
\begin{equation}
Z_{\pi^+ \to \mu^+ \to \nu_e} (\alpha) =
\frac{24\,\left( \alpha\,\left( 1 - r_\pi \right)  - 
      r_\pi\,\left( 1 - r_\pi^\alpha \right)  \right) }
    {\alpha^2\,\left( 1 + \alpha \right) \,
    \left( 2 + \alpha \right) \,
    \left( 3 + \alpha \right) \,
    {\left( 1 - r_\pi \right) }^2}
\end{equation}
with $r_\pi = (m_\mu/m_\pi)^2$.
The $Z$--factors for  charged conjugate modes
are  identical  because  of $CP$  invariance.
The  $Z$--factors given above 
are  calculated  neglecting the energy loss of muons  before
decay, and assuming  that the muon  helicity is exactly conserved.
This  assumption remains valid, to a very good approximation,
also in the presence of a  magnetic  field, because 
the  bending  of the momentum and the 
spin precession exactly cancel  for a particle
of electric charge $\pm 1$ and
 magnetic moment   of one~Bohr magneton.
Neglecting the effects of the muon polarization
leads to an  overestimate  (underestimate)
of  the $Z$--factor for the
 $\pi^+ \to \mu^+ \to \overline{\nu}_\mu$
($\pi^+ \to \mu^+ \to \nu_e$) channel.
 For  $\alpha = 2$  the $Z$--factors for pion decay 
take the  values:
\begin{eqnarray}
\{ 
Z_{\pi^+ \to \nu_\mu}, 
Z_{\pi^+ \to \mu^+ \to \overline{\nu}_\mu}, 
Z_{\pi^+ \to \mu^+ \to \nu_e}\}_{\alpha=2}
& = & \left \{ \frac{1-r_\pi}{2}, 
\frac{3 + 4 \,r_\pi}{20}, 
\frac{2 + r_\pi}{10} \right \}  \nonumber \\
& ~ &  \\
& \simeq & \{ 0.2135, ~0.2646, ~0.2573 \} ~, \nonumber
\end{eqnarray}
i.e. the three neutrinos  carry  approximately 
one quarter of the charged pion energy. This
happens  because  in the first decay  the muon
carries  away   a large fraction ($(1 + r_\pi)/2 \simeq 0.787$)
of the pion energy.

\begin{figure} [hbt]
\centerline{\psfig{figure=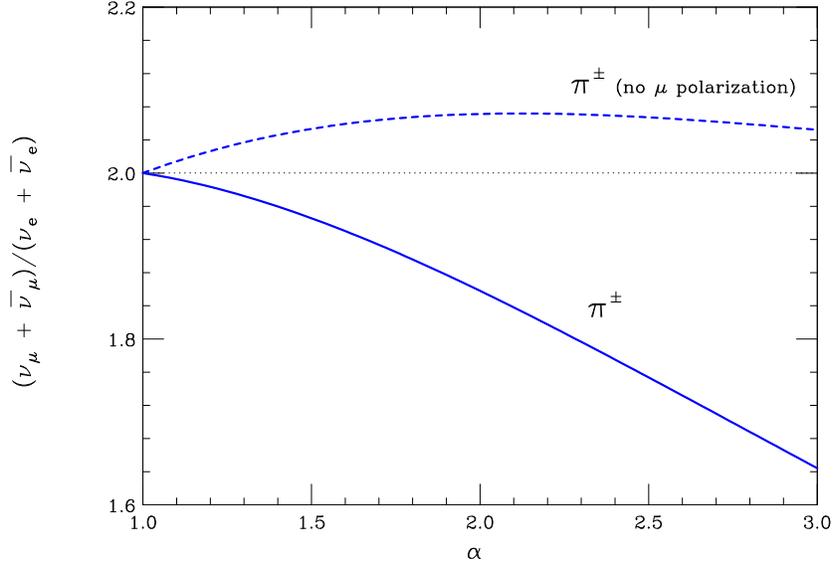,angle=90,width=11.0cm}}
\caption {
Neutrino flavor ratio  
 $(\nu_\mu + \overline{\nu}_\mu)/(\nu_e + \overline{\nu}_e)$
at the source  obtained from the chain decay of 
charged pions  with a  power law energy spectrum
$\propto E_\pi^{-\alpha}$.
Energy losses are assumed to be  negligible.
The dashed line  neglects the effect  of muon polarization
in pion decay.
\label{fig:z}  }
\end{figure}

Because of the different shapes of the energy distributions
of the three  neutrinos  emitted   in a pion decay,
the    flavor ratio
\begin{equation}
R_{\mu e} = \frac{(\nu_\mu + \overline{\nu}_\mu)}{(\nu_e + \overline{\nu}_e)}
= \frac{ Z_{\pi^+ \to \nu_\mu} + Z_{\pi^+ \to \mu^+ \to \overline{\nu}_\mu}}
{Z_{\pi^+ \to \mu^+ \to \nu_e}}
\end{equation}
for a power law  spectrum of parent pions
is a function of its  slope.
The   ratio 
is shown in  fig.~\ref{fig:z}.
For  $\alpha =1$  
one has  $R_{\mu e} = 2$, 
 since in this  case  the  $Z$ factors correspond
to the  neutrino multiplicities.
With increasing  $\alpha$   the flavor ratio decreases
monotonically,  this  reflects the fact  that the
muon neutrinos  produced in the direct pion
decay are softer  than the neutrinos  
of the  muon   decay. 
For the value  $\alpha=2$  one has   $R_{\mu e} \simeq 1.858$.
Figure~\ref{fig:z} also 
shows   the predictions obtained
neglecting the effects of 
muon polarization
to illustrate the importance of their inclusion.

\begin{figure} [hbt]
\centerline{\psfig{figure=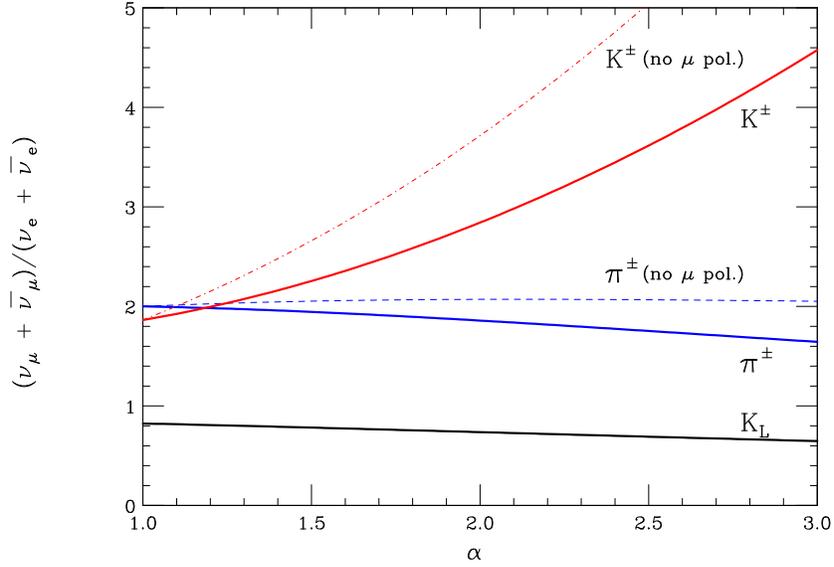,angle=90,width=11.0cm}}
\caption {
Neutrino flavor ratio    $(\nu_\mu + \overline{\nu}_\mu)/(\nu_e + \overline{\nu}_e)$
at the source   for   the chain decay of 
different  mesons
having  a  power law  spectrum $\propto E^{-\alpha}$.
The different lines refer  to the decay
of  $\pi^\pm$, $K^\pm$ and $K_L$. The dashed (dotdashed) line
  show the effect
of  neglecting muon polarization   in  the chain decay
of charged pions (kaons).
\label{fig:z1}  }
\end{figure}

The second most important source of  neutrinos  is the 
decay of  kaons. 
Charged kaons    can produce  neutrinos  with three
different  decay channels:
the     two body  mode
$K^+ \to \mu^+\nu_\mu$  
(with  branching ratio  0.6343), and  the 
three body modes
$K^\pm_{e3}$ 
($K^+ \to \pi^\circ e^+ \nu_e$)  and $K^\pm_{\mu3}$ 
($K^+ \to \pi^\circ  \mu^+ \nu_\mu$) 
that have branching ratios 
$B^\pm_{e3} \simeq 0.0487$
and  $B^\pm_{\mu3} \simeq 0.0327$  \cite{PDG}.
The $K_L$ can  also  produce neutrinos   
in the $K^\circ_{e3}$ decay mode 
($K_L \to  \pi^\mp e^\pm\nu_e (\overline{\nu}_e)$) with 
combined   branching ratio $B^0_{e3} \simeq 0.3881$,
and $K^\circ_{\mu3}$ mode 
($K_L \to  \pi^\mp \mu^\pm \nu_\mu (\overline{\nu}_\mu)$) with 
combined   branching ratio $B^0_{\mu3} \simeq 0.2719$.
The $R_{\mu e}$ ratios  for neutrinos  produced
by the chain  decay of   charged or neutral kaons
is  shown in fig.~\ref{fig:z1}. In the case of
charged kaons  the  2--body decay mode
is  the dominant one,
the $Z$--factors  for this  mode
have the same  form as in pion decay,
with the replacement  $r_\pi \to  r_K$,  however 
for  charged  kaons 
 the flavor  ratio  
grows  when $\alpha$ increases,
reflecting the  fact that the 
muon neutrinos   produced in the direct decay 
carry  nearly half ($(1-r_K)/2 \simeq 0.48$)
of the  parent particle 
 energy while the neutrinos  
produced in the second  stage (from muon decay)  take on average only 
18\% ($\nu_\mu$) and 16\% ($\nu_e$) of the kaon energy. 
The  inclusion  of the
three body decay modes 
($K^\pm_{e3}$
and $K^\pm_{\mu 3}$)  that have smaller branching fractions
reduces the  flavor ratio,   but it is a small
correction.

In the case of the decay of $K_L$,
the three body decays 
are the only  source of  neutrinos.
The $K^\circ_{e 3}$ mode  has a larger
branching  fraction  than the  $K^\circ_{\mu 3}$ channel,
 because of the larger  phase space  available. 
When  $\alpha = 1$, when 
the  flavor ratio  reflects the neutrino
multiplicities in the final  state,
the  flavor ratio  is 
$R_{\mu e} =2 B^\circ_{\mu 3}/( B^\circ_{e 3} + B^\circ_{\mu 3})
\simeq 0.82 $.
Increasing $\alpha$  the ratio   decreases,
because the  $\nu_e$ from the $K^\circ_{e 3}$ decay 
have the hardest spectrum,
and their contribution is   enhanced.

Note that    for power laws  with slope  $\simeq 2$ the  net effect
 of including kaon decay   as a  $\nu$ source  is  a 
only a small (positive) correction, because
of a  cancellation  between the  contributions 
of  charged  kaons (that  increase the  relative  importance
of muon neutrinos)  and neutral  kaons (that do the opposite).

\section{Energy Loss  Mechanisms}
\label{sec:pdec}
In the most general  case,
the unstable  particles  that create  neutrinos
can lose  a significant  amount  of energy  during the time
that  elapses  from  the moment of their creation
to the moment of their decay.
This effect  can have very important consequences  for the
spectrum and  flavor  composition of the neutrinos.

As an illustration,
muon energy losses are very important
for the prediction of the atmospheric  neutrino  fluxes.
The largest effect is  due to the fact that
the  muons  that  reach the ground  lose rapidly energy
because of ionization and electromagnetic radiation processes,  
and decay practically at rest (or are captured by heavy nuclei
\cite{Conversi:1947ig})
producing only  very low  energy neutrinos.
Pion  energy losses are
also very important for  the calculation of 
the  atmospheric  $\nu$ fluxes. In this 
case the dominant  mechanism for energy loss
are  hadronic  interactions in air.
The pion  decay length grows  linearly  with $E_\pi$, 
and   (for $E_\pi \gtrsim 100$~GeV)
it becomes  (for  vertical  particles)
comparable to the  interaction length.
Most pions of higher energy
interact before decaying.
These hadronic  interactions result in a multiplication
of  the pions,  but  the  secondary  particles have much lower
energy and the net  effect is a 
 strong  suppression of the 
neutrino  flux at high energy.  

There are  several mechanisms  for  energy loss
that could be present  in the  astrophysical 
environment where neutrinos are produced.
In several  circumstances synchrotron radiation 
could be the dominant source of  energy loss \cite{Rachen:1998fd,Kashti}.
In the presence of  a magnetic  field  of value $B$, 
the energy loss of a particle of electric charge $e$, mass $m$ and
energy $E$, after  averaging over all possible
orientation of the  magnetic  field  direction
is given by the well known expression:
\begin{equation}
-\frac{dE}{dt}  = \frac{4}{9} \frac{e^4 \, B^2}{m^4}\; E^2
\end{equation}
The energy loss  is   significant 
only for  $E \gtrsim E_{\rm syn}$
the critical energy where the 
the synchrotron loss time $t_{\rm syn} = -[(dE/dt)/E]^{-1}$ 
is  equal to the decay time $t_{\rm dec} = \tau \, E/m$:
\begin{equation}
E_{\rm syn} 
= \frac{3}{2} \frac{m^{5/2}}{e^2\, B \, \sqrt{\tau}} =
\frac{5.8 \times 10^{18}~{\rm eV}}{B_{\rm Gauss}}
 ~\left (\frac{m}{m_\mu} \right )^{\frac{5}{2}}
 ~\left (\frac{\tau}{\tau_\mu} \right )^{-\frac{1}{2}}
\label{eq:synloss}
\end{equation}
The critical energy for 
synchrotron losses scales  $\propto m^{5/2}\;\tau^{-1/2}$.
It is smallest for muons,
it becomes 18.4 times  larger  for  charged pions,
and  628  times  larger for charged kaons. 
There is  therefore  an  energy range where the losses
are only  significant for  muons,  and  a
second  energy range  where 
the synchroton losses
are significant for  both pions and muons  but not for kaons.

In  the presence  of  ordinary  matter in  gas form
the particles  can also lose energy because of
ionization and radiation processes
(with radiation dominating at large energies).
These energy  losses take the form
$-dE/dt \simeq \rho \,(a + b\, E)$ 
where $\rho$ is the medium density and  $a$  and $b$  
are slowly varying
coefficients  that depend on the gas composition.
For  a medium mostly composed of protons,
as it is likely in astrophysical  environments,
the critical energy for the energy losses
with  ordinary matter in gaseous form
(where the loss time  equals the
decay time)  is for muons:
\begin{equation}
E^\mu_{\rm matter} 
\simeq 6.5 \times 10^{18}~ (\rho_{-10})^{-1}~{\rm eV}
\label{eq:mugas}
\end{equation}
where  $\rho_{-10}$ is the  matter density  
in units of $10^{-10}$~g~cm$^{-3}$.

In the presence of a gas of ordinary matter
the dominant source of  energy loss
for  charged  pions  and  kaons   is due to  hadronic interactions.
The critical  energy 
at which the   hadronic  interaction  time
$t_{\rm int} 
=  \langle A \rangle /(N_A \, \rho \, \sigma_{int})$
equals the  decay  time $\tau \, E/m$ 
can   be estimated for pions as:
\begin{equation}
E_{\rm matter}^{\pi} \simeq 7.4 \times 10^{16}~
 ({\rho_{-10})^{-1}~{\rm eV}} ~.
\label{eq:pi-int}
\end{equation}
This  critical  energy scales   $\propto m/\tau$ and for
charged  kaons  (assuming the same  hadronic
cross section) is  approximately 7.4 times   larger.

The effects  of ordinary matter
on the secondary particles   can   be 
safely neglected for all energies 
(including those above  $E_{\rm matter}$) and all particle
types if  the   density is  below the critical  value:
\begin{equation}
\rho_{\rm crit} \simeq  2.0 \times 10^{-15} ~B_{\rm Gauss}~
{\rm g~cm}^{-3} ~.
\end{equation}
For  $\rho < \rho_{\rm crit}$ the  
energy  losses  on matter are   either  negligibly small, or
dominated  by the synchrotron losses.

\begin{figure} [hbt]
\centerline{\psfig{figure=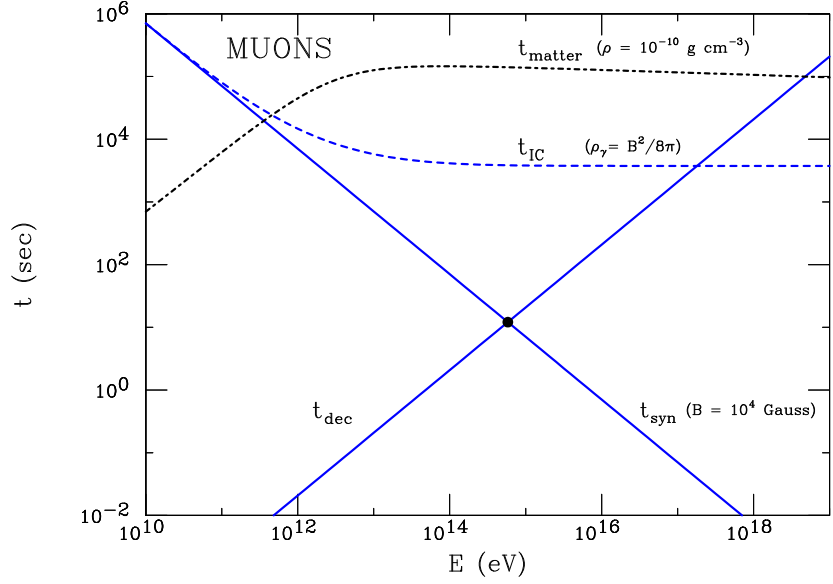,angle=90,width=11.0cm}}
\caption {
Characteristic  times  for a muon  plotted as a function 
of its energy.
The different lines  correspond to
(i) the
decay time $t_{\rm dec}$;
(ii) the time  $t_{\rm matter}$ for 
energy loss in ordinary matter,
assuming a gaseous medium with the average
composition of the interstellar medium (ISM)  and the  
density $\rho = 10^{-10}$~g~cm$^{-3}$
(this time scales as $t_{\rm matter} \propto \rho^{-1}$);
(iii) the time $t_{\rm  syn}$ for  synchrotron
losses  in a magnetic  field $B = 10^4$~Gauss
($t_{\rm syn} \propto B^{-1}$),
and  (iv) the time  $t_{IC}$ for  Inverse Compton  
losses.
The target radiation field is assumed  isotropic,
with energy spectrum 
of GRB form   (\protect\ref{eq:grb-spectrum})
with $\varepsilon_{\rm b} = 1$~KeV, and  energy density
$\rho_\gamma = \rho_B$  (with $B = 10^4$~Gauss).
\label{fig:time_mu}  }
\end{figure}

\begin{figure} [hbt]
\centerline{\psfig{figure=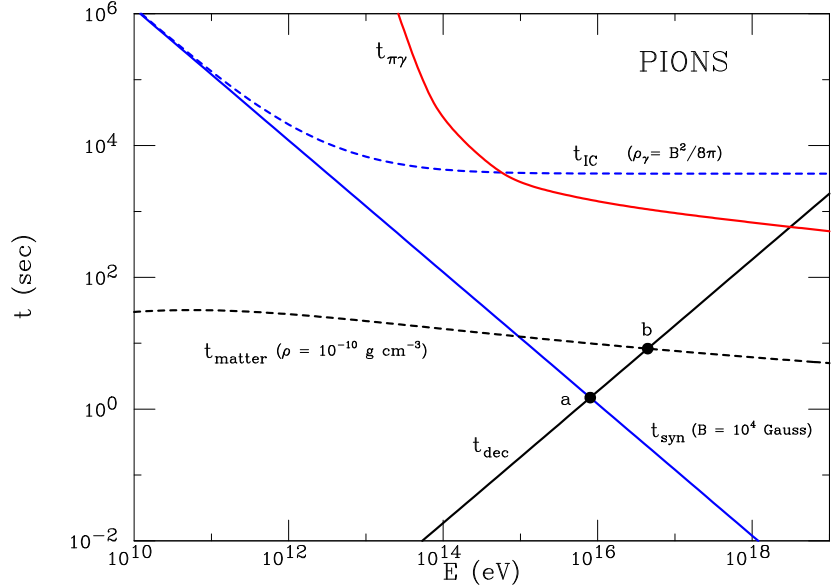,angle=90,width=11.0cm}}
\caption {
Characteristic  times  for charged  pions.
The different lines  correspond to: (i) the
decay time $t_{\rm dec}$;  
(ii) the time $t_{\rm matter}$ 
for hadronic interactions  with a  gaseous  medium 
with the ISM  composition and 
 density $\rho = 10^{-10}$~g~cm$^{-2}$;
(iii) the time $t_{\rm syn}$ for synchrotron  losses
(in a randomly oriented  magnetic  field $B = 10^4$~Gauss);
(iv)  the time $t_{IC}$ for 
Inverse Compton losses;
(v) the time $t_{\pi\gamma}$ for 
photo--hadronic  interactions.
The  last two times  are
 calculated for a 
radiation field of GRB form   (\protect\ref{eq:grb-spectrum})
with $\varepsilon_{\rm b} = 1$~KeV, and energy density
$\rho_\gamma = \rho_B$ (with $B = 10^4$~Gauss).
Point a  (b) 
is the intersection  of 
the $t_{\rm dec}$ and  $t_{\rm syn}$  ($t_{\rm matter}$) curves.
The  energy that correspond to the  point
is the critical energy above which synchrotron losses
(the hadronic interaction probability)   are (is)  important 
before decay.
For the  values  of $B$ and $\rho$   chosen in this  illustration
the presence of a gas of ordinary matter is  negligible.
\label{fig:time_pion}  }
\end{figure}

Figures~\ref{fig:time_mu}
and~\ref{fig:time_pion}
illustrate  the above discussion
showing the characteristic times  $t_{\rm dec}$ (decay),
$t_{\rm syn}$  (synchrotron losses)  and  $t_{\rm matter}$ 
for muons and pions.

Gamma Ray  Bursts have been proposed
by Waxman and Bahcall 
\cite{Waxman-grb}   
as a neutrino  source
having a potentially very strong magnetic  field.
In the WB model, the strength of the magnetic  field
is  estimated   by energy equipartition,
assuming that  it  corresponds to an energy density
of the same  order of magnitude as 
the  energy density carried by the fireball photons.
It is   natural to ask the question if  this radiation field
can be  an important  source of energy loss
for secondary particles produced in it.
The
 calculation
of the time $t_{IC}$  for Inverse Compton losses
of charged  particles   traveling in 
the fireball radiation field
  requires to take into account
the energy spectrum of the target photons 
(assumed to be isotropic in the GRB jet frame).
In fact  the Compton scatterings with sufficiently high energy
photons $\varepsilon \gtrsim  m^2/E$
happen in the   Klein--Nishina  regime   and  are inefficient
as  a source of  energy loss.
For the predicted 
GRB photon spectrum \cite{Waxman-grb},
(given  also  in 
  equation (\ref{eq:grb-spectrum}))
that  falls at high energy $\propto \varepsilon^{-2}$, 
the  energy loss   for Inverse Compton
 grows at high energy 
   as $-(dE/dt)_{IC} \propto E \, \rho_\gamma/m^2$,
(where $\rho_\gamma$ is  the   target photon energy density),
and the  loss time  $t_{IC}$  becomes  constant.
The   results of a detailed  integration
are  also  shown in fig.~\ref{fig:time_mu}
and~\ref{fig:time_pion}, where  we have assumed
that  the densities $\rho_B = B^2/(8 \pi)$ and
$\rho_\gamma$ are equal.  
In  most circumstances (unless equipartition is 
very badly violated)  Inverse Compton  losses
can be neglected.

Charged  pions (and kaons) traveling in a radiation 
field can also lose  energy for photo--hadronic
interactions. For  completeness we have calculated 
the interaction  time of  pions in the  
GRB radiation field. The result is also shown in 
fig.~\ref{fig:time_pion}.
For this  calculation the $\pi \gamma$   hadronic cross section 
has  been   estimated as the  sum 
of  contributions for production of the 
 relevant    resonances \cite{PDG} :
$\rho(770)$, $a1(1260)$, $b1(1235)$ and $a2(1320)$ 
and a  non resonant background.
At high energy   ($s \gtrsim 30$~GeV$^2$)
the cross  section is   described 
by a  formula obtained using Pomeron universality and Regge poles 
factorization from the fits to the $\gamma$-proton and 
$\pi$-proton cross sections reported in \cite{PDG}.
The lower  energy  non--resonant  background
has  been    normalized to obtain a smooth energy
dependence.

At very high energy pion--photon interactions
are more important that Inverse Compton losses,
however  if the energy density in
photons and magnetic  field are comparable,
the synchrotron losses dominate. 

\subsection{Decay Energy Distribution}
To describe the effect of energy losses,
 it is  useful
to consider the  function  
$dp_{\rm dec} /dE (E, E_i)$ that  gives the 
probability   density  for an
unstable  particle  of  mass $m$ and lifetime $\tau$,
 created  with initial  energy $E_i$
to  decay   with energy  
$E$. 
Assuming that the energy loss  is a continuous  process,
and  can be described  by the equation:
\begin{equation}
-\frac{dE}{dt} (E) = f (E) ~,
\end{equation}
(with  $f(E)$ is the average loss per unit time),
the  distribution
$dp_{\rm dec} /dE (E, E_i)$ 
can be   calculated   as:
\begin{equation}
\frac{dp_{\rm dec}}{dE} (E; E_i) =
\frac{m}{\tau} \; \frac{1}{f(E) \; E} 
~\exp \left [-\frac{m}{\tau} \; \int_{E}^{E_i} 
\frac{dE^\prime} {f(E^\prime) \, E^\prime} \right ].
\end{equation}
For an  energy loss of  
form $f(E) = a \, E^n$  (with $n > 0$)
the  decay probability becomes \cite{Kashti}:
\begin{equation}
\frac{dp_{\rm dec}}{dE} (E; E_i) = \frac{\epsilon^n}{E^{n+1}} ~
\exp \left [-
\frac{\epsilon^n}{n} \; \left ( 
\frac{1}{E^n} -
\frac{1}{E_i^n}
 \right )\right ]
\end{equation}
with:
\begin{equation}
\epsilon = \left (  \frac{m}{\tau \, a} \right )^{\frac{1}{n}}.
\end{equation}
The quantity $\epsilon$  has  the physical  meaning of the energy for
which  the  loss-time:
$t_{\rm loss} = E/f(E) = 1/(a \, E^{n-1})$ 
is  equal to the decay time:
$t_{\rm decay} = \tau \, E/m$.

\begin{figure} [hbt]
\centerline{\psfig{figure=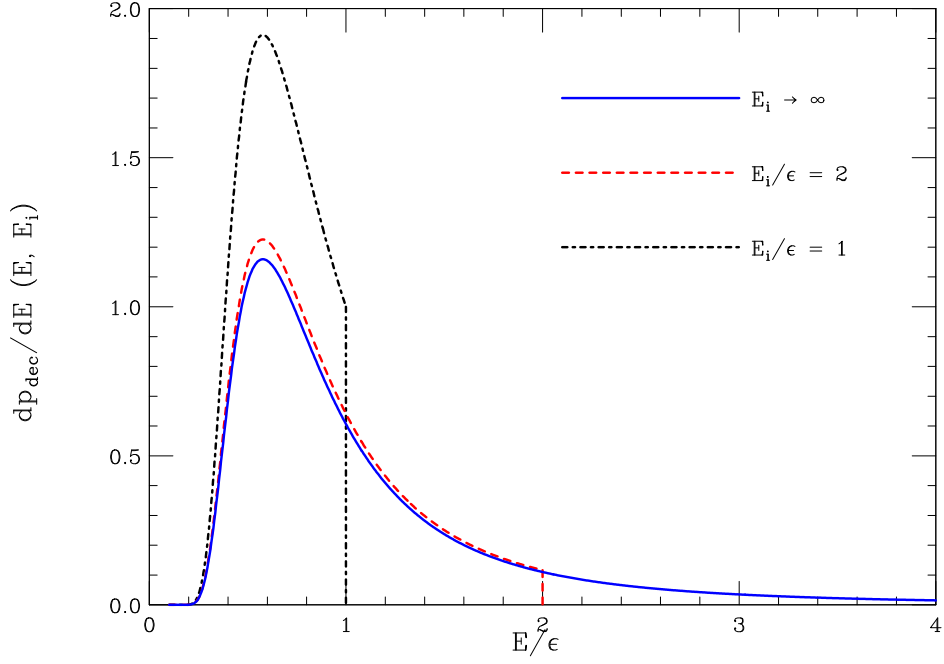,angle=90,width=12.5cm}}
\caption {
Decay energy distribution $dp_{\rm dec}/dE (E, E_i)$  for  particles
losing energy as $-dE/dt = a\,  E^{2}$. The critical energy 
$\epsilon = \sqrt{m/(a\tau)}$ 
 is the energy where  the  decay time
and the energy loss time are equal.
\label{fig:prob}  }
\end{figure}

For  $E_i \ll  \epsilon$, the energy loss
is    negligible and 
 the decay   energy   distribution  becomes
a simple delta function:
\begin{equation}
\left (
\frac{dp_{\rm dec}}{dE}  (E; E_i) 
\right )_{E_i \ll \epsilon}
  \simeq  \delta [E - E_i],
\end{equation}
while for  $E_i \gg \epsilon$ 
the  decay energy distribution takes a universal  form
\begin{equation}
\left (
\frac{dp_{\rm dec}}{dE}  (E; E_i)
\right )_{E_i \gg \epsilon}
 = \frac{\epsilon^n}{E^{n+1}} ~
\exp \left [-
\frac{1}{n} 
\frac{\epsilon^n}{E^n} 
\right ].
\end{equation}
An illustration  of the  decay energy distribution
for the case $f(E) = a  \;E^2$, relevant for 
synchrotron emission is  shown in  fig.~\ref{fig:prob}.
The important  qualitative   feature  in  fig.~\ref{fig:prob}
is that all particles  with   sufficiently  high initial
energy   ($ E_i \gtrsim  2~\epsilon$)
  have  nearly identical   final  energy   distributions,
with a well defined  maximum  at $E \sim 0.6 \, \epsilon$.
In many  cases  this  can result
in ``pile--up'' effects in the final state particle spectra.

\section{Hadronic  interactions}
\label{sec:inter}
The    target  for the primary particles
 interactions   can be either 
normal matter in gaseous  form, or a radiation  field.
In the following we will consider both cases  separately.

\subsection{Gas Target}
For a  target material    made of normal matter at rest
the interaction  rate  of a primary particle
of energy $E_p$  is:  
\begin{equation}
K_{pp} (E_p) = n_{\rm gas} ~\sigma_{pp} (s).
\label{eq:pgas}
\end{equation}
where $n_{\rm gas}$  is the  number   density
of the target material.
Equation (\ref{eq:pgas}) is  valid  for 
a situation 
where  the   primary particles  are protons and
the target  material is  also dominated by protons.
This is  likely to  be a good approximation in most
circumstances.   It is straightforward to
consider the  more   general  case.

Since the target is at rest,
the interaction rate is   proportional
to the cross section at a well defined
c.m. energy  $s = m_p^2 + 2 m_p \, E_p$.
Since  hadronic  cross sections grow  only 
logarithmically  with   c.m. energy  (or as  power law  
$\propto  s^\epsilon$ with  a small exponent $\epsilon$), the
interaction  rate of the  primary  particles  
changes  only very slowly  with energy.

It is well known   that the   multiplicity
and  energy distributions of the  particles produced in 
hadronic  interactions    cannot  be calculated from
first  principles, however   it is  believed that 
particle  production, to a  a reasonably good approximation,
satisfies  Feynman scaling, defined  by the condition:
\begin{equation}
E \; {d\sigma_j  \over dp_\parallel^*}  (p_\parallel^*, \sqrt{s}) =
 F_j(x_F)
\label{eq:fey-scaling}
\end{equation}
where $p_\parallel^*$  is  the longitudinal momentum  in the
c.m. frame,
$d\sigma_j$ is the inclusive differential cross section 
for the production of particle type  $j$, 
and
$x_F =  2 \, p_\parallel^*/ \sqrt{s}$
is the Feynman  variable.

For  large c.m.  energy ($s \gg m_p^2$)   the  
target  rest frame  energy   of secondary  particles
in the forward  hemisphere ($x_F > 0$) is  well approximated
by  the expression $E \simeq E_p \; x_F$
(where $E$  ($E_p$)  is the energy of
the final  state  (projectile) particle in
this frame). It follows  
that the validity of Feynman--scaling 
in the fragmentation region
implies  also the approximate validity 
of  scaling  of  the inclusive  cross  sections 
in the target  rest  frame:
\begin{equation}
{dn_{pp \to a}  \over dE} (E; E_p) 
\simeq {1 \over E_p} ~
F_{pp \to a}\left  ( {E \over E_p} \right )
~~~~~~ ({\rm for} ~~ E \gg m)
\label{eq:z-scaling}
\end{equation}
The   approximate validity of the scaling law (\ref{eq:z-scaling})
together  with the slow  variation 
of  the  hadronic cross sections  with c.m. energy  have the
important consequence
(based on   essentially the same  argument previously
discussed in connection with decay (\ref{eq:zz}))  that
a  power law  spectrum
of primary particles   generates  spectra of secondaries
also having  a power law form.
For  example for  final state particles of  type $\pi$:
\begin{eqnarray}
Q_\pi (E_\pi) & = & 
C_p \; n_{\rm gas} \; \sigma_{pp} 
\;  E_\pi^{-\alpha} ~\int_0^1 ~dx ~x^{\alpha -1} 
~ F_{pp \to \pi}(x) 
= C_p \; n_{\rm gas} \; \sigma_{pp} \;
Z_{pp \to \pi} (\alpha) \; E_\pi^{-\alpha} ~.
\label{eq:zz1}
\end{eqnarray}

In  general, one  expects  some  deviations
from a perfect  power law.
First of all, this can reflect  a shape of
the primary spectrum  different  from a
power law. In addition  the 
energy dependence of the cross sections, and 
the existence  of   violations of  Feynman
scaling (as  measured  in   hadron  colliders)
introduce  some distortions.
Another situation   that  can  result
in   large  deviations from a power law  emission
arises  when  particles  of different
rigidities  have different  confinement volumes  having
different  target  densities (while our derivation of 
(\ref{eq:zz1}) implicitely assumed a   homogeneous
source  volume).

\subsection{Photoproduction}
In  several  of the proposed  neutrino  sources,
the target of the primary particles is a radiation field.
In  this  case  the  interaction  rate and the energy
distribution of the particles  produced in an
interaction depend   not only on the density, but also
on the energy and  angular distribution  of the
target photons. 
The  interaction probability  per unit time 
 of a proton of energy $E_p$
traveling in  the  radiation 
field  described  by  
$n_\gamma (\varepsilon, \Omega_\gamma)$ can be calculated as:
\begin{equation}
K_{p\gamma} \left (E_p \right ) 
= \frac{1}{\lambda_{p\gamma} (E_p)} =
\int d\varepsilon  ~
\int_{-1}^{+1}  \; \frac{d\cos \theta_{p\gamma}}{2}  \; 
(1-\cos\theta_{p\gamma}) ~ 
n_\gamma (\varepsilon, \cos \theta_{p \gamma}) ~
\sigma_{p \gamma} (\epsilon_r)
\label{eq:kpgam0}
\end{equation}
where $\theta_{p\gamma}$ is the 
angle  between the  photon   and the proton  momenta 
in the interaction,
$\sigma_{p \gamma}$ is the
photoproduction   cross section, and 
$\epsilon_r$ is the   photon energy   in the proton rest  frame:
\begin{equation}
\epsilon_r = \frac{E_p \, \varepsilon}{m_p}
\; (1 - \cos \theta_{p\gamma})
\end{equation}
The quantity $\epsilon_r$ is in one to one correspondence
with the c.m. energy  of the reaction.
It is   convenient to  change  the   integration 
 variable  from $\cos\theta_{p\gamma}$
to $\epsilon_r$. Restricting
ourselves  to a  situation (and a frame)
where  the photon distribution
is  isotropic  one can then rewrite 
(\ref{eq:kpgam0}) as:
\begin{equation}
K_{p\gamma} (E_p) =  \frac{1}{2} \, \frac{m_p^2}{E_p^2} \;
~\int_{\epsilon_{\rm th}}^\infty \;  d\epsilon_r ~\epsilon_r 
 ~ \sigma_{p \gamma} (\epsilon_r)
\int_{(m_p \epsilon_r)/(2 E_p)}^\infty
\;  d\varepsilon~\frac {n_\gamma(\varepsilon)}{\varepsilon^2} 
\label{eq:kpgam}
\end{equation}
where $\epsilon_{\rm th}$ is the  threshold  photon energy
for pion production in the proton rest frame:
\begin{equation}
\epsilon_{\rm th} = m_\pi + \frac{m_\pi^2}{2 m_p}
\end{equation}
Equation (\ref{eq:kpgam})  can be recast  in the  form
\begin{equation}
K_{p\gamma} (E_p) =  
~\int_{\epsilon_{\rm th}}^\infty \;  d\epsilon_r ~
f_{p\gamma} \left (\epsilon_r; ~E_p\right ) ~.
\label{eq:kpgam1}
\end{equation}
This  expression   shows  explicitely the fact
that   the $p\gamma$ interactions  of a proton
of energy $E_p$  
do not correspond to a single
value of the c.m. energy but  have a distribution, that
in general  is a function of  $E_p$, and is  determined  by the
energy (and angular) distribution of the   target photons.
The  probability distribution for  $\epsilon_r$  is:
\begin{equation}
p (\epsilon_r; ~E_p) =
  \frac{f (\epsilon_r, E_p)}{K_{p \gamma} (E_p)} ~.
\end{equation}

A phenomenologically important case    for the 
target  radiation field  is the form:
$n_\gamma (\varepsilon) = C_\gamma \, \varepsilon^{-\beta}$,
that is  an (isotropic) 
 power law  spectrum  of  slope $\beta$.
For this form  
the last  integration in (\ref{eq:kpgam})  can be  
performed analytically   with the result:
\begin{equation}
K_{p\gamma} (E_p) =  C_\gamma \;
\frac{2^\beta}{\beta+1}
\; \left (
\frac{E_p}{m_p} \right )^{\beta-1} ~
\int_{\epsilon_{\rm th}}^{\infty} ~
d\epsilon_r ~   \epsilon_r^{-\beta}
~ \sigma_{p \gamma} (\epsilon_r)
= K_0 (\beta) ~E_p^{\beta -1}
\label{eq:kpgam2}
\end{equation}
One can see that the interaction rate   has the energy dependence 
$E_p^{\beta-1}$. 
The  growth with energy (for $\beta > 1$) of the 
interaction rate  can be understood 
observing  that 
a proton of  energy $E_p$ can interact inelastically
only  with photons  above  a minimum  energy 
$m_p \, \epsilon_{\rm th}/(2 E_p)$.  This  minimum
target energy  decreases proportionally to  $E_p^{-1}$, 
and therefore protons  of  higher  energy can interact  with 
a softer and more abundant  photon population.
A result   essentially equivalent to
equation (\ref{eq:kpgam2}) was  originally shown   by 
Waxman and Bahcall in  \cite{Waxman-grb}.
An important  feature of equation (\ref{eq:kpgam2})
is that
the  probability distribution for $\epsilon_r$
takes a form that is independent from $E_p$:
\begin{equation}
p (\epsilon_r) =
  \epsilon_r^{-\beta}
~ \sigma_{p \gamma} (\epsilon_r) 
\left [\int_{\epsilon_{\rm th}}^{\infty} ~
d\epsilon_r ~ 
  \epsilon_r^{-\beta}
~ \sigma_{p \gamma} (\epsilon_r) 
\right ]^{-1}
\label{eq:pr}
\end{equation}
This   fact  has  interesting consequences  discussed below.
It should be noted  however that 
equations 
(\ref{eq:kpgam2})
and (\ref{eq:pr})
  are
calculated  assuming   a power law  that  extends
to very large energy  without any cutoff.
This is   not  only unrealistic,
but  in   some important cases also
untenable.
In fact 
the  integral  over  $\epsilon_r$
in  (\ref{eq:kpgam2}) 
diverges at its  upper  limit
 (since
$\sigma_{p\gamma}$ is   slowly increasing  with  energy)
 for  $\beta \le 1$, and the expression  for $K_{p \gamma}$ becomes
meaningless.
The    divergence   corresponds to the divergence
of the  photon  number  density 
in the absence of an upper limit  cutoff
(note that in  fact the energy  density of the photon  population
diverges  at the upper  end  already for   $\beta \le  2$). 
The introduction of a  high energy cutoff
for the energy distribution of the target photons 
is  therefore  mandatory.  This will be discussed 
in the next section.

The  properties  of particle production in
$p\gamma$ interactions  is clearly  intimately related
to the distribution of c.m. energy (or equivalently
$\epsilon_r$)   of the interactions.
It is necessary to consider the energy distributions
of  secondary particles in the ``source frame''
(where the   primary proton has  energy $E_p$). In this frame 
the energy of a secondary particle 
can obviously be expressed as function of
quantities in the c.m.  frame   with an appropriate
Lorentz boost:
\begin{equation}
E = \gamma \; ( E^* +  v \, p_z^*)
\end{equation}
where $v$  and $\gamma$ are the velocity 
and gamma--factor of the c.m. of the reaction in the
source frame, $E^*$  is the c.m. energy of the
secondary particle  and $p_z^*$  the momentum  component
parallel to $\beta$.
The Lorentz $\gamma$    that connects the source
and c.m.  frames is:
\begin{equation}
\gamma = \frac{E_p + \varepsilon}{\sqrt{s}} \simeq
\frac{E_p}{\sqrt{s}}  ~.
\end{equation}
In the second approximated  equality we have neglected
the photon energy $\varepsilon$ with respect
to $E_p$,   this is 
expected to be an excellent  approximation.
Similarly one can safely  make the approximation
$v \simeq 1$.
With these approximations 
we can rewrite the Lorentz  boost
from the c.m. to the source frame 
as:
\begin{equation}
E =
 E_p \;   \frac{E^* + p_z^*}{\sqrt{s}} =
 E_p \; \xi  ~.
\end{equation}
This equations    indicates  that to a good approximation
all   secondary particles of  source frame  energy $E$ 
are created in the c.m.  frame of  the interaction with  
(to a very good approximation)  the same value
of the quantity $\xi = (E^* + p_z^*)/\sqrt{s}$.
The energy spectrum of  (for example) pions
created in the interaction of a proton
of energy $E_p$  can then be   written as:
\begin{equation}
\frac{dn_{p\gamma \to \pi}}{dE_\pi}  (E_\pi; ~E_p) \simeq
\frac{1}{E_p} ~ \int_{\epsilon_{\rm th}}^\infty
d\epsilon_r ~ p(\epsilon_r ;E_p) ~ 
\left [ \frac{dn_{p \gamma \to \pi}}{d\xi}
(\xi;~\epsilon_r)
\right ]_{\xi = E_\pi/E_p}
\label{eq:pgamma1}
\end{equation}
where  $dn_{p\gamma \to \pi}/d\xi (\xi, \epsilon_r)$ 
is the $\xi$ distributions  of the secondary  particles of type
$\pi$ in $p\gamma$ interactions with c.m. energy
that correspond to $\epsilon_r$;  this  distribution
is  convoluted for 
a  fixed  value   $\xi = E_\pi/E_p$  over all possible
value of the  $\epsilon_r$  
with the appropriate  distribution.
If  the 
 energy  distribution of the target photon field has a power law  
form, the    function  $p(\epsilon_r;~E_p)$ is independent
from $E_p$,  and  equation (\ref{eq:pgamma1}) becomes
the expression of a scaling law of  form:
\begin{equation}
\frac{dn_{p\gamma \to \pi}}{dE_\pi}  (E_\pi; ~E_p) \simeq
\frac{1}{E_p} 
~ F_{p\gamma \to \pi}
 \left ( 
\frac{E_\pi}{E_p};~\beta
 \right )
~.
\label{eq:pgamma2}
\end{equation}
The scaling function $F_{p\gamma \to \pi}$   is not
truly universal, but depends on the 
slope of the   target photon spectrum.
Integrating  over all   primary particle energies one can obtain
the production rate of  pions as:
 \begin{eqnarray}
Q_\pi (E_\pi) & = & 
\int_{E_\pi}^{\infty} ~dE_p ~N_p (E_p) ~ K_{p \gamma} (E_p) 
~{dn_{p\gamma  \to \pi} \over dE_\pi} (E_\pi ; E_p) \nonumber  \\
& = & 
\int_{E_\pi}^{\infty} ~dE_p ~\left (C_p \; E_p^{-\alpha} \right )~
~ \left ( K_0 (\beta) \; E_p^{\beta-1} \right )
{1 \over E_p} ~
 F_{p\gamma  \to \pi}\left  ( \frac{E_\pi}{ E_p}; ~\beta \right )
\nonumber \\
& = & 
C_p \; K_0 (\beta)  E_\pi^{-(\alpha-\beta + 1)}
 ~\int_0^1 ~d\xi ~\xi^{\alpha  -\beta} 
~ F_{p\gamma \to \pi}(\xi;~\beta)
\nonumber \\ 
& =  & C_p \;  K_0 (\beta) 
\; Z_{p\gamma  \to \pi}[\alpha-\beta+1,\beta]
 \; E_\pi^{-(\alpha-\beta+1)}
~~.
\label{eq:zz2}
\end{eqnarray}
This  equation   shows  that  under 
the assumptions  made, that   is:
(i)  a power law spectrum  of protons, and
(ii) an isotropic power law  spectrum of target photons,
the  energy distribution of  produced  secondaries
is  again a power law with slope $\alpha -\beta +1$.
Note that this result
is  based on purely kinematical considerations, 
while the   result (\ref{eq:zz1})
about the  interactions on  an ordinary matter target
was  based on a dynamical  assumption  about the 
approximate  validity of Feynman  scaling  in hadronic
interactions, and  on the   weak   energy 
dependence  of  hadronic  cross sections.

A  result  roughly  equivalent to (\ref{eq:zz2})   was
obtained   by Waxman and Bahcall in   \cite{Waxman-grb},
who concluded  that in the case
$\beta = 1$   the neutrino  emission
has a power law  spectrum
with the same slope
 of the parent  protons.
However,  the  validity  of equation (\ref{eq:zz2}),
as  the validity of   equation (\ref{eq:kpgam2}),
 relies on the untenable assumption  that the  power law   spectrum
of   target photons   has no  high energy cutoff.
The  important effects of the existence of an high energy cutoff
for  equations   (\ref{eq:kpgam2}) and (\ref{eq:zz2})
will be discussed in section~\ref{sec:grb}.

\subsection{Nuclear  photodisintegration}
The beta  decay of neutrons   produces  $\overline{\nu}_e$. In
many  circumstances   the contribution
of this  anti--neutrino source   is  of negligible importance.
This is  because  the  average 
fraction of the neutron energy  $E_n$ carried away 
by the   $\overline{\nu}_e$ 
(in a frame where the neutron is  ultrarelativistic)
 is 
$\langle E_{\overline{\nu}_e}\rangle/E_n \simeq 5.1 \times 10^{-4}$,
with end--point 
$E^{\rm max}_{\overline{\nu}_e}/E_n \simeq  2 (m_n - m_p - m_e)/m_n
\simeq 1.66 \times 10^{-3}$.
In most cases  the   flux of   the softer $n$--decay neutrinos 
is  neglible  with respect to the contribution
of  neutrinos from $\pi/K$ of decay  that   carry a much larger
fraction (of order $\sim 0.25$) of their parent energy.
There are   two circumstances  where
the $n$ contribution can become  important.
The  first one  is when  the neutrino spectrum    has a
low energy cutoff.  This  happens   naturally when
the  target of  the primary particles is a radiation field
and there is  an interaction   energy  threshold.
In this  case the $\overline{\nu}_e$   produced in neutron
decay can become the dominant
component  of the neutrino flux  at low  energy, because
most of the neutrinos from pion and kaon decay  have higher  energy.
An example of this situation will be   shown in the following
section.
A second, more interesting   case is  when the   neutrons are produced
in the photodisintegration  of  high  energy nuclei
\cite{neutron-decay}.
In this  case it is  in principle possible to
have   the emission of a pure  $\overline{\nu}_e$ flux.
In fact the threshold for photodisintegration
of a nucleus of mass number $A$ expressed in terms
of energy per nucleon
($E_0 = E_{\rm tot}/A$) is  of order:
\begin{equation}
\left (E_0 \right )_{\rm th}^{\gamma A}
 \simeq \frac{ m_p \epsilon_{\rm bind}}{2 \; \varepsilon_\gamma}
\label{eq:photodisintegration}
\end{equation}
where $\varepsilon_\gamma$ is the  energy of the target photons
and $\epsilon_{\rm bind} \simeq 8$~MeV is the binding energy
 of a  nucleon  in the nucleus. 
The threshold for pion photoproduction  is: 
\begin{equation}
\left (E_0 \right )_{\rm th}^{\pi}
 \simeq \frac{ m_p  \; m_\pi}{2 \; \varepsilon_\gamma}
\left (1 + \frac{m_\pi^2}{2 \, m_p} \right )  ~.
\label{eq:photopion}
\end{equation}
Since  the binding energy $\epsilon_{\rm bind}$ 
is approximately  fifteen  times  smaller
than a pion mass, it is in principle
possible to   have circumstances where the  primary particles 
are below   the threshold for  pion production, but above the threshold
for  photodisintegration.
This  would clearly   result  in a  pure $\overline{\nu}_e$ flux.
It should be noted   that 
the  photodisintegration and pion photoproduction thresholds differ
by only one order of magnitude, and therefore
a pure  $\overline{\nu}_e$  flux  can only extend 
for a   small interval of  energy. 
In general one expects important contributions 
from neutrinos from meson decays.
In fact, even a very small   number of 
pion  production  interactions 
created by the high energy tails
 of  the photon and/or    primary particles spectra
can  result in an important  ``contamination'' 
of   $\nu_e$ and  muon neutrinos.  Most of these neutrinos
have an energy two orders of magnitude
larger  that the $\overline{\nu}_e$ produced in $n$ decay
and  have  a correspondingly larger  cross section.

\section{Neutrino Production in Gamma Ray Bursts}
\label{sec:grb}
As an  illustration of  the  spectra  and 
flavor composition   of neutrinos
from astrophysical sources,
in this section we will  consider  the 
model  of neutrino  production  in Gamma  Ray Bursts  (GRB)
developed by Waxman and  Bahcall  (WB) \cite{Waxman-grb}.
Our goal  is not to analyze  critically the assumptions
that underline the  model,
but to recalculate
 with a more detailed  modeling of  particle production
the neutrino fluxes starting  from  the same general assumptions.

In the  (WB)   model  \cite{Waxman-grb}
 the GRB  prompt  emission 
is  generated by the  synchrotron radiation 
of  high energy electrons   accelerated 
by internal  shocks in a  relativistic  expanding wind
(for  more discussion  see  \cite{Piran-grb,zhang-meszaros};
alternative  explanations of the  GRB mechanism
exist in the literature, see for  example \cite{dar-derujula-grb}).

The internal shocks also accelerate  protons, that can
interact  with the radiation field  of the wind
and produce  secondary particles  
that  generate neutrinos.
The accelerated protons   have a power law 
spectrum of  slope  $\alpha \simeq 2$,  while
the target radiation field
is isotropic  in the  wind frame,
and approximated as a broken power law of form:
\begin{equation}
n_\gamma(\varepsilon) = \cases {
C_\gamma \; \varepsilon^{-1}  
&  for $\varepsilon  \le \varepsilon_{\rm b}$, \cr
C_\gamma \; \varepsilon_{\rm b} \;\varepsilon^{-2}  
&  for $ \varepsilon > \varepsilon_{\rm b}$
}
\label{eq:grb-spectrum}
\end{equation}
where   $\varepsilon_{\rm b}$ is a   break energy.
The   photon  energy distribution 
in the wind frame is related to the 
obervable  spectrum of the GRB prompt   emission
by  a Lorentz   transformation with parameter  $\Gamma \simeq 300$.
Since the observed  break energies are  distributed around
an average value
 $\langle \varepsilon_{\rm b}^{\rm obs} \rangle \sim 300$~KeV
 \cite{batse}, 
the typical  value  of  $\varepsilon_{\rm b}  $
in the wind frame is of order 1~KeV.

The energy spectra
of   individual  GRB 
are indeed  well fitted  by  two  power laws
smoothly joined  at a  break energy, 
however the  fitted values of 
the  exponents  of the spectrum   have  rather  broad
distributions    centered at
$\beta \simeq 1$ below the break energy, 
and
$\beta \simeq 2.2$ above the break energy \cite{batse}.
In our  calculation we have therefore
considered a   generalization of  (\ref{eq:grb-spectrum})
that  leaves the  two exponents as  free parameters.
We have also  introduced
a  high energy cutoff
$\varepsilon_{\rm max}$, because
in its absence  the energy
density of the radiation field  would  diverge
for a high energy slope $\le 2$.
Our description of the  energy spectrum of the 
target radiation field  then  becomes:
\begin{equation}
n_\gamma(\varepsilon) = \cases {
(C_\gamma/\varepsilon_{\rm b}) \;
 (\varepsilon/\varepsilon_{\rm b})^{-\beta_1}  
&  for $\varepsilon  \le \varepsilon_{\rm b}$, \cr
(C_\gamma/\varepsilon_{\rm b}) \;
 (\varepsilon/\varepsilon_{\rm b})^{-\beta_2}  
&  for $ \varepsilon_{\rm b} < \varepsilon  <
 \varepsilon_{\rm max}$, \cr
0   &  for $ \varepsilon  \ge 
 \varepsilon_{\rm max}$. \cr
}
\label{eq:grb-spectrum0}
\end{equation}
The calculation of the interaction rate 
 of a proton of energy $E_p$ in 
such a radiation field  is  straightforward.
The target photon field (\ref{eq:grb-spectrum0}) 
can be  seen as  the sum of two  power law
spectra   with    sharp  low and high--energy
cutoffs
$\varepsilon_{\rm min}$ and $\varepsilon_{\rm max}$.

We have already  obtained the expression for
the proton interaction  rate for 
a power law   that  extends to all energies.
The expression  (\ref{eq:kpgam2}) 
must however  be modified  in the presence
of low and high energy cutoffs.
In the presence of a high energy cutoff
$\varepsilon_{\rm max}$
the  interaction  rate    vanishes  below a 
threshold  energy $E_{\rm th}$.
Above  the threshold,  in the interval:
\begin{equation}
E_{\rm th} =  \frac{m_p \; \epsilon_{\rm th}} 
{2 \, \varepsilon_{\rm max} } \le 
 E_p \le 
 \frac{m_p \; \epsilon_{\rm th}} 
{2 \, \varepsilon_{\rm min} }.
\end{equation} 
 the   interaction rate  is
\begin{equation}
K_{p\gamma} (E_p) = 
\frac{2^\beta \, C_\gamma }{\beta+1}\; \left (
\frac{E_p}{m_p} \right )^{\beta-1} ~
\int_{\epsilon_{\rm th}}^{2 E_p \varepsilon_{\rm max}/m_p} ~
d\epsilon_r ~ 
 \sigma_{p \gamma} (\epsilon_r) ~
\left [  \epsilon_r^{-\beta} - 
\epsilon_r \, \left ( 
\frac {m_p}{2 E_p \, \varepsilon_{\rm max}}\right )^{\beta+1}
\right ]
\label{eq:kpgam3}
\end{equation}
coinciding with (\ref{eq:kpgam2}) in the limit
$\varepsilon_{\rm max} \to \infty$.
The effect of  a lower energy cutoff $\varepsilon_{\rm min}$
is to  stop the $E_p^{\beta-1}$  growth of 
the interaction  rate.
For  $E_p >  m_p \epsilon_{\rm th}/(2 \varepsilon_{\rm min})$ 
 equation (\ref{eq:kpgam3})  must be  replaced by:
\begin{eqnarray}
K_{p\gamma} (E_p) & = & 
\frac{C_\gamma \, m_p^2 }{2 \, E_p^2 \, (1 + \beta)} ~
(\varepsilon_{\rm min}^{-(\beta + 1)} -
\varepsilon_{\rm max}^{-(\beta + 1)}) ~
 ~ \int_{\epsilon_{\rm th}}^{2 E_p \, \varepsilon_{\rm min}/m_p} ~
d\epsilon_r ~ \epsilon_r ~
 \sigma_{p \gamma} (\epsilon_r) 
\nonumber \\
& ~ &  ~ \nonumber \\
& + & 
\frac{2^\beta \, C_\gamma}{\beta+1}\; \left (
\frac{E_p}{m_p} \right )^{\beta-1}
\int_{2 E_p \varepsilon_{\rm min}/m_p}^{2 E_p 
\varepsilon_{\rm max}/m_p} ~
d\epsilon_r ~ 
 \sigma_{p \gamma} (\epsilon_r) ~
\left [  \epsilon_r^{-\beta} - 
\epsilon_r \, \left ( 
\frac {m_p}{2 E_p \, \varepsilon_{\rm max}}\right )^{\beta+1}
\right ]  ~~~~
\label{eq:kpgam4}
\end{eqnarray}

For  a qualitative understanding,  it can  be 
useful to   consider 
the $p\gamma$
cross  section as approximately constant
above  threshold.
The  integrations
over  $\epsilon_r$ in (\ref{eq:kpgam3}) and (\ref{eq:kpgam4})
are then    trivial.
For  $\beta \ne  1$   one finds:
\begin{equation}
\left [
K_{p\gamma} (E_p) 
\right ]^{\beta \ne 1} 
\simeq \cases{  0  & for $x < 1$ \cr
C_\gamma \, \sigma_{p \gamma} ~ 
\left (
\frac {2 \, \varepsilon_{\rm max}^{1-\beta}}{\beta^2 - 1} \right )
~\left [ x^{\beta -1} - 
\frac{1}{2} \left ( 1 + \beta + (1-\beta) \, x^{-2} \right )
\right ]
 & for $   1 \le x \le  r$ \cr
C_\gamma \, \sigma_{p \gamma} ~ 
\varepsilon_{\rm min}^{1-\beta} ~
\left [ \frac{1 - r^{1-\beta}}{\beta -1} -
\frac{1}{x^2 \, (1 + \beta) } \;
\left (r^2 - r^{1-\beta} \right ) 
\right ] & for $    x >  r$
}
\label{eq:kcorr1}
\end{equation}
where  $x = E_p/E_{\rm th}$ 
is the proton energy expressed  in units 
of the threshold  energy, and
  $r = \varepsilon_{\rm max}/\varepsilon_{\rm min}$.
For   $\beta = 1$, one  finds:
\begin{equation}
\left [K_{p\gamma} (E_p)
\right ]^{\beta = 1} 
 \simeq \cases{  0  & for $x < 1$ \cr
C_\gamma \, \sigma_{p \gamma} ~ \left [ \log[x] - \frac{1}{2} 
\, \left ( 1 - \frac{1}{x^2} \right )
\right ]
 & for $   1 \le x \le  r$ \cr
C_\gamma \, \sigma_{p \gamma} ~ \left [ \log[r] - \frac{1}{2 \, x^2} 
\, \left ( r^2 - 1\right )
\right ] & for $    x >  r$ \cr
}
\label{eq:kcorr2}
\end{equation}
These expressions   show that the
interaction  rate  for  a power law
spectrum of target photons  in 
a restricted range  do maintain   the 
approximate  behaviour $K_{p\gamma} \propto E_p^{\beta -1}$ 
but only  in  a  limited  energy  region
(in the limit $\beta \to 1$ the 
interaction rate      grows  logarithmically: 
$K_{p \gamma} \propto  \log[E_p \, \varepsilon_{\rm max} ]$).
The interaction probability  vanishes below  threshold
and    goes asymptotically to a constant for
very large   $E_p$.
It can also be useful to consider
the    high energy limit  ($E_p/E_{\rm th} \to \infty$)
of expressions (\ref{eq:kcorr1}) and (\ref{eq:kcorr2}):
\begin{equation}
\left [K_{p \gamma} (E_p) \right ]^{\beta \ne 1}_{E_p \to \infty}
 = \sigma_{p \gamma} \;
\frac{C_\gamma}{\beta-1} ~
\left (
\varepsilon_{\rm min}^{-\beta +1}  
-\varepsilon_{\rm max}^{-\beta +1}  
\right )  ~,
\end{equation}
\begin{equation}
\left [K_{p \gamma} (E_p) \right ]^{\beta = 1}_{E_p \to \infty}
=  \sigma_{p \gamma} \;
C_\gamma
~\log \left [
\frac{\varepsilon_{\rm max}}{\varepsilon_{\rm min}}
\right ]~.
\end{equation}
These  expressions  have the form
$K_{p \gamma} = \sigma_{p \gamma} \; N_\gamma$ 
with $N_{\gamma}$ the integrated  number  density
of the target photons,  that can be immediately
recognized as   the correct  high energy limit.

The  interaction probability   for the  radiation field
(\ref{eq:grb-spectrum0})  can be obtained  combining
two  expressions corresponding to the 
parts  of the photon spectrum below and above 
the  break energy $\varepsilon_{\rm b}$ that plays
in  the two cases the role
of the maximum  or the minimum   target photon energy. 
An illustration  of the  energy dependence of the
proton interaction rate 
is  shown in  figure~\ref{fig:k_pgamma}.
The curves in this  and the following   figures
are made independent from the value of 
$\varepsilon_{\rm b}$ measuring all   energies in units  of:
\begin{equation}
E^* 
= \frac{m_p \, \epsilon_{\rm th}}{2\,  \varepsilon_{\rm b}}
= \frac{m_p  \, m_\pi }{2 \,\varepsilon_{\rm b}}
\; \left ( 1 + \frac{m_\pi} {2 \, m_p} \right ) 
\simeq 6.9 \times 10^{13}
~\left (
 \frac{\varepsilon_{\rm b}}{\rm KeV} \right )^{-1}
~{\rm eV} .
\label{eq:estar}
\end{equation}
The energy $E^*$ has  the physical meaning
of the  threshold proton energy for
inelastic  interactions with photons  having
the break  energy $\varepsilon_{\rm b}$.
In figure~\ref{fig:k_pgamma} 
we   have chosen for
the    target photons energy spectrum the original 
 form (\ref{eq:grb-spectrum})
with  exponents $\beta_1 = 1$, $\beta_2 = 2$ and
no  high energy cutoff.
For low  $E_p$ the interaction rate  
grows approximately  linearly with energy
according to the scaling law (\ref{eq:kpgam2})
because   low energy protons
interact inelastically  only   with the  high energy part
of the  radiation field  (with slope $\beta_2 = 2$).
For  larger  energy ($E_p \gtrsim E^*$)   
the growth becomes  logarithmic  following 
the qualitative behaviour of  equation (\ref{eq:kcorr2}).
It should be noted  that 
the assumption made in \cite{Waxman-grb}
 of considering 
the  interaction rate as  approximately constant for $E_p > E^*$ 
is  not a good  approximation.

\begin{figure} [hbt]
\centerline{\psfig{figure=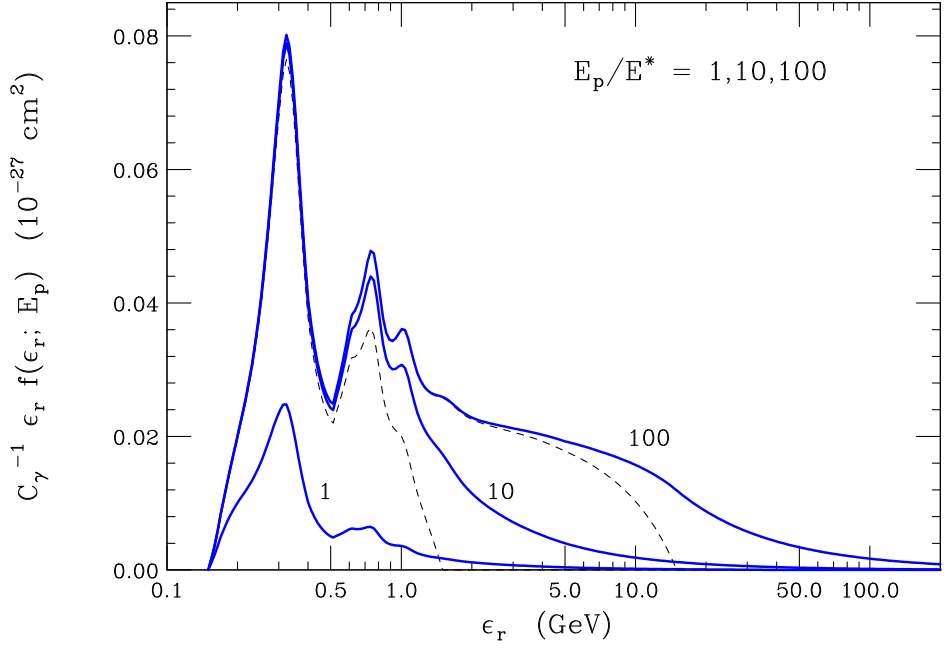,angle=90,width=12.5cm}}
\caption {
Distribution  of $\epsilon_{\rm r}$ 
(the   photon  energy in  the  proton 
rest  frame) 
for the  $p\gamma$ interactions of  protons of  different 
energy $E_p$.
The isotropic  photon field  has the 
energy spectrum  of GRB form  (\protect\ref{eq:grb-spectrum}).
The dashed line  neglect  the contribution of photons
above  the break energy $\varepsilon_{\rm b}$.
\label{fig:er_dist}  }
\end{figure}

In fig.~\ref{fig:er_dist} 
we show the distribution  $f(\epsilon_r; ~E_p)$ 
of the c.m.  energy of the interaction for
different  values of  the proton energy
($E_p/E^* =1$, 10, 100). 
The distribution  is  proportional
to the  $p\gamma$ cross section, and the  shape 
of the curves  in the figure  
reflects the energy dependence of $\sigma_{p\gamma}$ 
that has prominent resonances. 
The    most important one  (at $\epsilon_r \sim 0.3$~GeV)
corresponds to the production of the $\Delta$ resonance.
The important  feature in the figure 
is that with increasing energy, higher and higher c.m.  energies
become  possible.
This  has  significant  phenomenological consequences.
For example,  for $\epsilon_r$  sufficiently small,
only production of a single  pion,
in the two channels
$p \pi^\circ$ and $n\pi^+$,
 is kinematically  allowed,
 and therefore
there is no  $\pi^-$  production, and  consequently
no $\overline{\nu}_e$ production.
For larger c.m.  energy  multiple  pion   production
becomes possible,  $\overline{\nu}_e$ can be produced, 
and the ratio  $\nu_e/\overline{\nu}_e$  decreases. 
Above the  $K \Lambda$  energy threshold,
also the production of kaons  becomes possible
introducing an additional  neutrino  source.

In the following we will show   some  examples of the
neutrino fluxes   that are obtained
varying the parameters  of the model.
Our  calculations  
have been   performed   by Montecarlo  methods   using a detailed
model for the $p\gamma$ cross section  that includes 
the production of all relevant resonances and a non--resonant 
component \cite{photo-mc}. 
In order to  perform a calculation
one needs to specify: (i)  the primary proton
spectrum, (ii) the target photon  spectrum, and
finally the value of the magnetic  field $B$ (all other 
sources of energy losses for  secondary particles  are
considered as negligible).
The   target photon energy distribution  
is taken with the form
(\ref{eq:grb-spectrum0})
that  depends on  4 parameters:
\{$C_\gamma$, $\varepsilon_{\rm b}$, $\varepsilon_{\rm max}$,
$\beta_1$, $\beta_2$\}.  Different  values
of $\varepsilon_{\rm b}$   correspond   to  different
values for  $E^* \propto \varepsilon_{\rm b}^{-1}$.
The proton spectrum  is taken  with the form:
\begin{equation}
N_p (E_p) = C_p  \; E_p^{-\alpha} ~\exp \left [ -
\left ( \frac{E_p} {E_{\rm max}}\right )^2 \right ]
\end{equation}
that is a power law  with a smooth  high energy cutoff:
defined  by  the three parameters:
\{$C_p$, $\alpha$, $E_{\rm max}$\}.
In all of the following we have kept  fixed
two of the parameters, the  proton maximum energy:
$E_{\rm max}/E^* = 10^4$, and the photon  high energy
cutoff $\varepsilon_{\rm max}/\varepsilon_{\rm b} = 300$.
The parameters   $C_p$ and  $C_\gamma$  determine
the  absolute value of the   neutrino emission,  but not 
its shape, and   will be  left unspecified
in the following.

The  absolute  value   of the neutrino flux
from an individual source 
can be obtained  specifying the values
of $C_p$ (that fixes the amount of energy 
in relativistic  protons  in the wind frame),
$C_\gamma$ that  specifies the  density of 
the photon target field, the exact  Lorentz Boost
and  obviously also the   source redshift  (distance).  
The integration over  the ensemble
of all sources  requires  additional  assumptions  about the
distributions of the     relevant parameters 
used for their   description  and  of their cosmological evolution.
In this  work  we will not discuss this  integration.

The value $B$ of the magnetic  field
is  in a one to one correspondence
with the muon  synchrotron energy, 
defined in equation (\ref{eq:synloss}), 
that can also be expressed as:
\begin{equation}
\epsilon_\mu =
\frac{E^\mu _{\rm syn}}{E^*} = 8.4 \times 10^4 
~\left (
\frac{ \rm Gauss}{B} 
\right )
~\left (
 \frac{\varepsilon_{\rm b}}{\rm KeV} \right )
\label{eq:epsmu}
\end{equation}

\begin{figure} [hbt]
\centerline{\psfig{figure=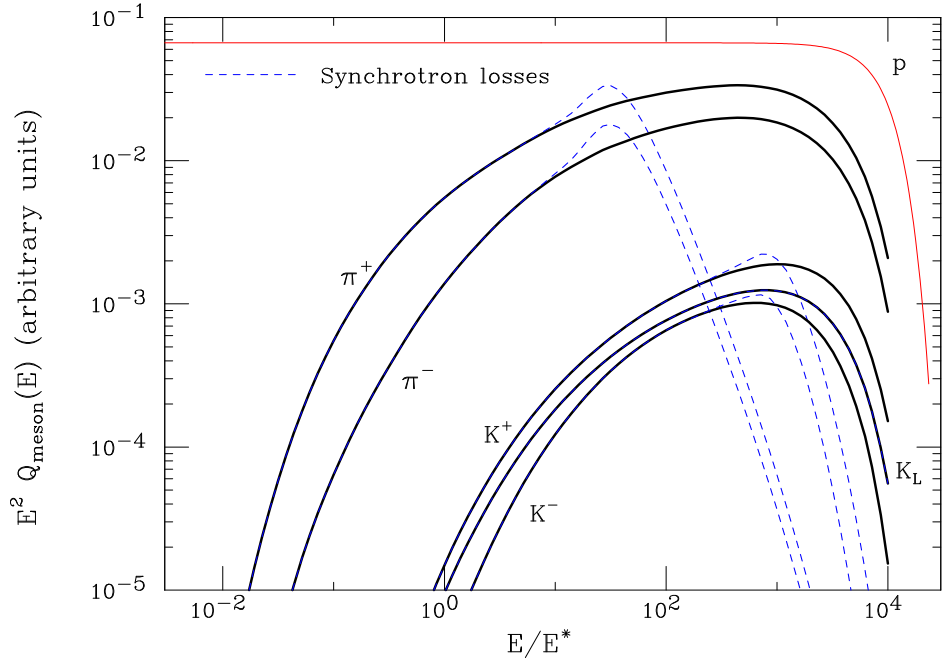,angle=90,width=12.5cm}}
\caption {
Energy distributions 
(in the form $Q_j (E) \, E^2 $ versus  $E$)
for  different  secondary  particles
($\pi^\pm$, $K^\pm$, $K_L$)
produced  by the  $p\gamma$ interactions 
of relativistic protons  in a GRB  radiation field.
The shape of the primary proton spectrum 
is  also shown 
(in the form $N_p (E) \, E^2$ versus  $E$)  
as a thin solid line.
The   energy spectrum of the isotropic
 target photons  has the form (\protect\ref{eq:grb-spectrum}).
All energies are  in units of  the
characteristic  energy
 $E^*$    (defined in equation (\protect\ref{eq:estar})).
The dashed  lines   are  the    distributions of the
meson  energy at  decay  assuming the presence of 
magnetic  field  with value 
$B = 2.78 \times 10^3~(\varepsilon_{\rm b}/{\rm KeV})$~Gauss.
\label{fig:meson}  
}
\end{figure}

As  an intermediate step   toward the calculation of
the neutrino fluxes 
we  show in  fig.~\ref{fig:meson}   the   
yields of  different  mesons. 
The   meson  yields  are calculated
for a proton spectrum   with  exponent $\alpha =2 $,
and  a radiation field   with slopes
(below and above the break energy)
 $\beta_1 = 1$ and $\beta_2  = 2$. 
The solid  lines  in fig.~\ref{fig:meson} 
show the energy distributions 
of  the  different  mesons at the moment of their creation.
In the absence of significant energy losses
the curves also  describe the  energy  distributions
at decay.
One   can see that   all  possible mesons  are produced,
but with different   abundances.
The most  abundant particles are 
charged pions, 
reflecting the fact  that a  large  fraction
of  the $p\gamma$ interactions   happens  close  to threshold
where only  single  (neutral or positive) pion  production is possible.
The production of $\pi^-$ 
becomes possible only above the threshold
for two  pion production, and is therefore  suppressed.
The production of  kaons   is   significantly smaller,
because it is   suppressed  both 
 dynamically  (since  it requires the creation of
 $s\overline{s}$ pairs)  and kinematically
(because of the larger strange hadron  masses).
The   threshold for the
production  of   mesons  containing
a strange  antiquark $\overline{s}$  ($K^+$ and $K^\circ$) 
is  lower,   since  it
corresponds to  final  states containing a strange 
baryon   (such as  $\Lambda \, K^+$). 
The production of  
mesons   containing a strange  quark $s$
 ($K^-$ and $\overline{K}^\circ$)  has a  higher threshold
since the   final state must contain  a  minimum of two kaons
(as for  example:  $p K^+ K^-$). Accordingly one finds
that the productions of  different kaon types   is  ordered 
as follows:  $K^+ > K_L > K^-$.
The   shapes of   energy spectra  of the mesons  
are  never  well approximated by a simple  power law,
but are  always ``curving'' in  a  log--log   representation.
The shapes of these spectra    reflect  the energy dependence
of the proton  interaction rate, and the 
opening up of the different  kinematical  channels.
At the highest  energy the   rapid  drop in the   meson yields 
is  connected to the  cutoff in the primary proton energy
at $E_{\rm max}/E^* \sim 10^4$.
Note that the ratio  $\pi^-/\pi^+$ 
and $K/\pi$ are not constant  but increase  with energy 
reflecting  the increasing  importance
of $\pi^-$ and kaon  production with  growing  c.m. energy.

The dashed  lines  in fig.~\ref{fig:meson} 
describe the decay energy distributions of the 
produced  mesons
assuming the   presence of  a magnetic field
that corresponds  to $\epsilon_\mu = 3$
(or   $B = 2.8 \times 10^3 ~
\varepsilon_{\rm b}^{\rm KeV}$~Gauss).
Pions  created  with energy
above their  critical  energy for
synchrotron losses
(in units of $E^*$: 
 $\epsilon_\pi \simeq  18.4  \epsilon_\mu \simeq 55$)
lose  most of their energy before  decay, 
and therefore the number of  pions  decaying
above this critical energy is 
strongly suppressed 
($\propto  (E_\pi/\epsilon_\pi)^{-2}$).
One can also notice that  synchrotron  losses
result in an enhancement of the 
 number of  pions    decaying  with   energy  just below 
$\epsilon_\pi$.  This   is 
a pile--up  effect   due to the fact that 
all  high energy pions  (with $E_\pi \gg \epsilon_\pi$)
decay with  similar energy distributions.
The same  effects  are present also for charged kaons
at higher energy, because the 
synchrotron  energy  for  charged kaons  (in unit of $E^*$) is 
$\epsilon_K \simeq 628\epsilon_\mu \simeq 34 \epsilon_\pi \simeq
1900$.

Secondary neutrons  are also  produced
in $p\gamma$ interactions. 
Assuming  that they can freely exit from  the source,
the  $\overline{\nu}_e$  spectra
from their decay have  been calculated 
and included in the following figures.

\begin{figure} [hbt]
\centerline{\psfig{figure=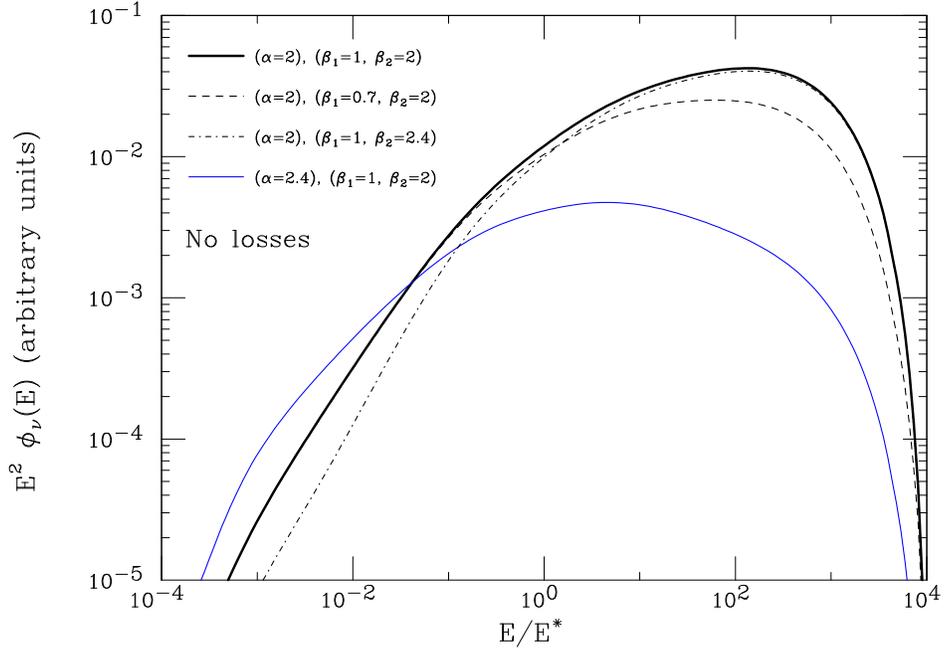,angle=90,width=12.5cm}}
\caption {
Neutrino energy spectrum (summed over all flavors) from
a GRB source,  according to the model \protect\cite{Waxman-grb}.
The calculation is performed   assuming a proton  spectrum
$N_p (E) \propto E^{-\alpha}$, traveling  in an  isotropic 
radiation field of form 
(\protect\ref{eq:grb-spectrum}). The different curves
are  obtained  varying the exponent $\alpha$ of the 
proton  spectrum, or the exponents $\beta_1$ and $\beta_2$ of the 
radiation  field. All  energies are measured  as  fraction
of $E^* =  m_p \, \epsilon_{\rm th}/(2 \varepsilon_{\rm b})$.
The energy losses of  all secondary  particles are
assumed to be negligible.
\label{fig:nu1}  }
\end{figure}

\begin{figure} [hbt]
\centerline{\psfig{figure=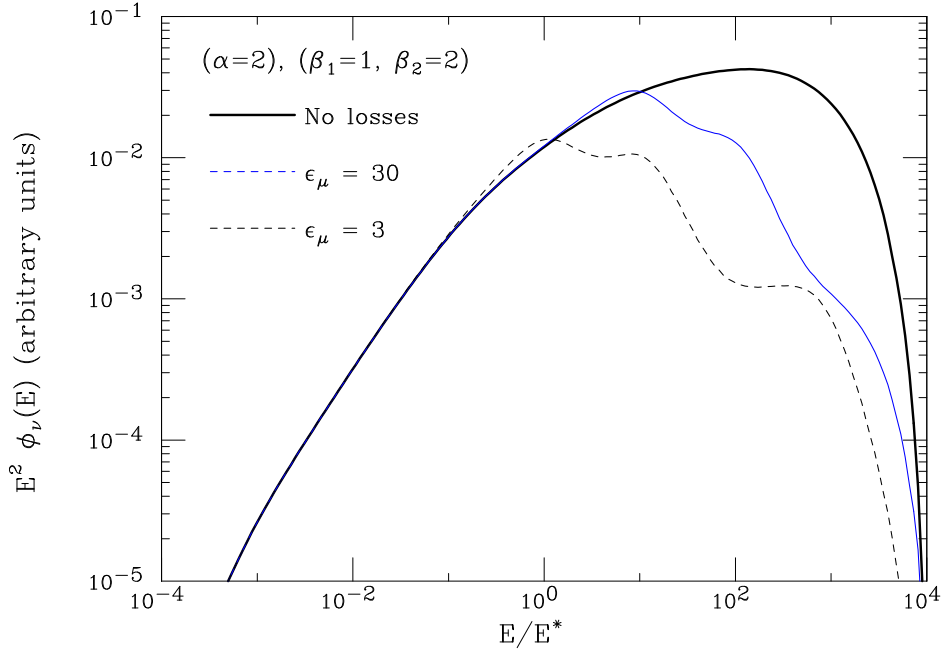,angle=90,width=12.5cm}}
\caption {
Neutrino energy spectrum (summed over all flavors) from
a GRB source (see text).  The  primary proton spectrum
has slope $\alpha = 2$,  while the target photons
have a broken power law energy distribution
(see equation (\ref{eq:grb-spectrum})) 
with slopes $\beta_1 = 1$ and $\beta_2 = 2$ below and above the
break  energy $\varepsilon_{\rm b}$.
The curves are calculated   assuming
different  values for the magnetic  field,
corresponding to values 
of the  critical energy for  muons
$\epsilon_\mu = \infty$, 30 and 3
(see eq. (\protect\ref{eq:epsmu})).
\label{fig:nu1a}  }
\end{figure}

Some examples of the resulting neutrino  fluxes
obtained  summing over all   possible parent
particles and all  $\nu$  types 
are  shown in 
fig.~\ref{fig:nu1} and fig.~\ref{fig:nu1a}.
The lines in fig.~\ref{fig:nu1}
are  calculated assuming  that energy losses for  secondary particles
are negligible. 
The  different  curves
correspond to different assumptions
about  the shape  of  the energy spectra for the primary protons
and target photons.
The thick solid  curve  is  calculated
with  the choice of  slopes 
(for the proton  flux and for the radiation field
below and above the break  energy):
$\{\alpha, \beta_1, \beta_2\} = \{2,1,2\}$.
The other  three curves show  the effect 
of changing the slopes one by one.
The largest  effect  is  related to the slope
of the primary protons,
increasing $\alpha$ from
2 to 2.4  results  in an important
softening of the neutrino  spectrum.
The modification of  the shape of the energy distribution 
of the target  photons also  distorts
the   neutrino spectrum.
Changing  the slope 
below    (above)  the break  energy,
modifies  the high  (low) energy 
part of the neutrino  spectrum.
This  is easily understood, since
high  (low)  energy neutrinos  are 
produced   by 
 higher (lower)  energy protons  that mostly interact
with  lower (higher)  energy photons.
Note that (reflecting the spectra of the parent
pions and kaons) the neutrino  energy distribution
changes gradually its slope
and  cannot be well fitted by a power law.

Figure~\ref{fig:nu1a}  illustrates   the 
effect  of including   synchrotron  losses
on  the    neutrino fluxes.
The  different  lines in  fig.~\ref{fig:nu1a}
show  the  neutrino
flux (summed over all  $\nu$  types) 
calculated  with the slopes
$\{\alpha, \beta_1 , \beta_2\} = \{2,1,2\}$,
and three  different assumptions for
the magnetic field  that correspond
to muon synchrotron energy $\epsilon_\mu = \infty$, 30 and 3
(the first case corresponds to negligible losses).
Increasing  the magnetic field  (reducing $\epsilon_\mu$)
suppresses the  neutrino flux at high energy.
In general, in the presence of  a strong 
magnetic  field  in the source, one can identify
four  interesting  neutrino energy ranges,
that are related to the ordering of the  synchrotron
critical energies for different  particles:
 $\epsilon_\mu  <  \epsilon_\pi < \epsilon_K$.
At sufficiently  low  energies  the synchrotron  losses
are completely negligible;
at higher  energy   the muons energy losses  have  to be taken
into  account;  at still higher energy
also the losses of charged  pions   must be considered;
at the  highest energy   the  emission
of synchrotron  radiation  is important also for charged kaons.
In general, in the presence of important
synchrotron emission,  one can 
have some  pile--up effects,   due to the fact
that all  high energy particles
decay just below their critical  synchrotron energy.

\begin{figure} [hbt]
\centerline{\psfig{figure=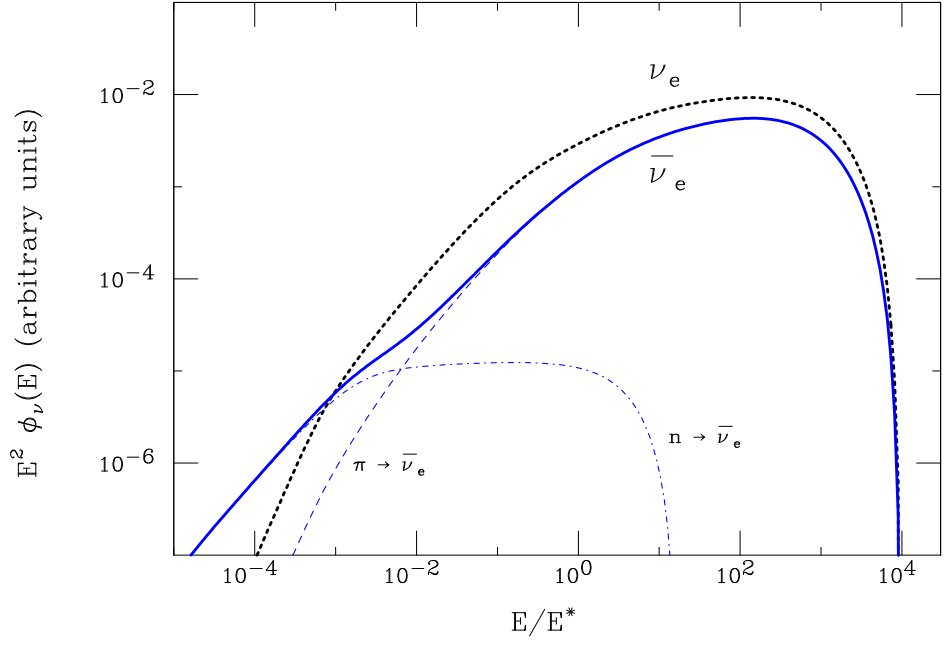,angle=90,width=12.5cm}}
\caption {
Energy spectra for
$\nu_e$ and $\overline{\nu}_e$ 
from a GRB source
(proton slope $\alpha = 2$,  target photon
slopes $\beta_1 = 1$ and $\beta_2 = 2$, no magnetic field)
For the  $\overline{\nu}_e$ flux also the  contributions
from   meson and neutron  decay  are separately shown.
Neutron  decay dominates at low energy.
\label{fig:neutron}  }
\end{figure}

The inclusive neutrino spectra shown in the previous  figures
contain  the  contribution of
the $\overline{\nu}_e$ from $n$ decay.  The structure
of this  contribution is  illustrated  in fig.~\ref{fig:neutron}.
The  anti--neutrinos  from neutron decay  have  significantly
lower  energies  that  the  neutrinos  from meson decay.
They  are the main component of the neutrino
flux  at the lowest energies,  where  however
they are difficult to observe because  the flux
is  suppressed  and the cross section is small.

\begin{figure} [hbt]
\centerline{\psfig{figure=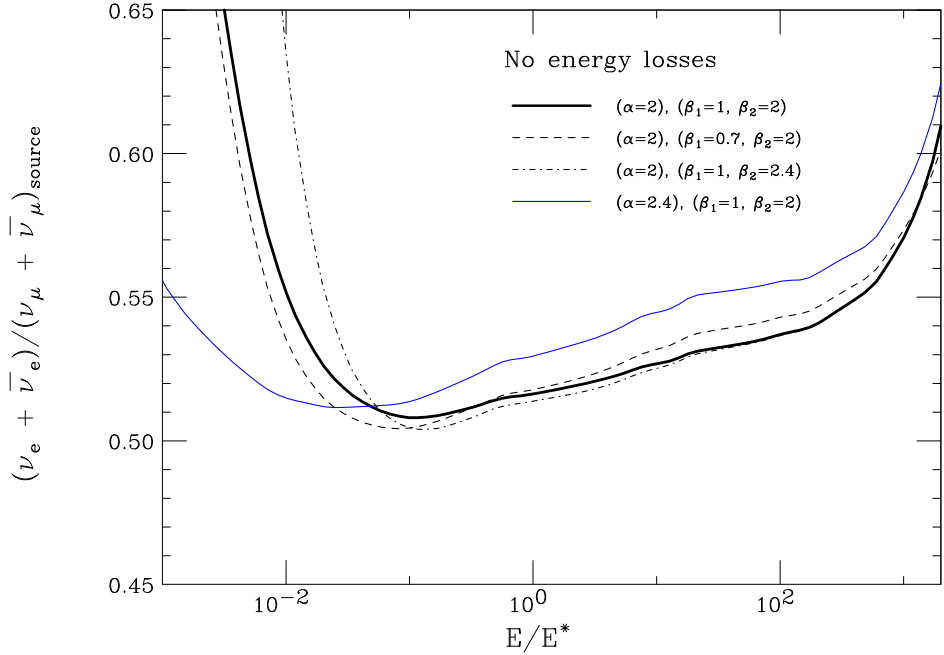,angle=90,width=12.5cm}}
\caption {
Ratio $(\nu_\mu + \overline{\nu}_\mu)/(\nu_e + \overline{\nu}_e)$  for 
neutrinos from a  GRB source. The  ratio   correspond
to the spectra shown in  fig.~\protect\ref{fig:nu1}.
The  rapid increase at low  energy  
is due to the  effect  of $\overline{\nu}_e$ from $n$ decay.  
\label{fig:nu2}  }
\end{figure}

\begin{figure} [hbt]
\centerline{\psfig{figure=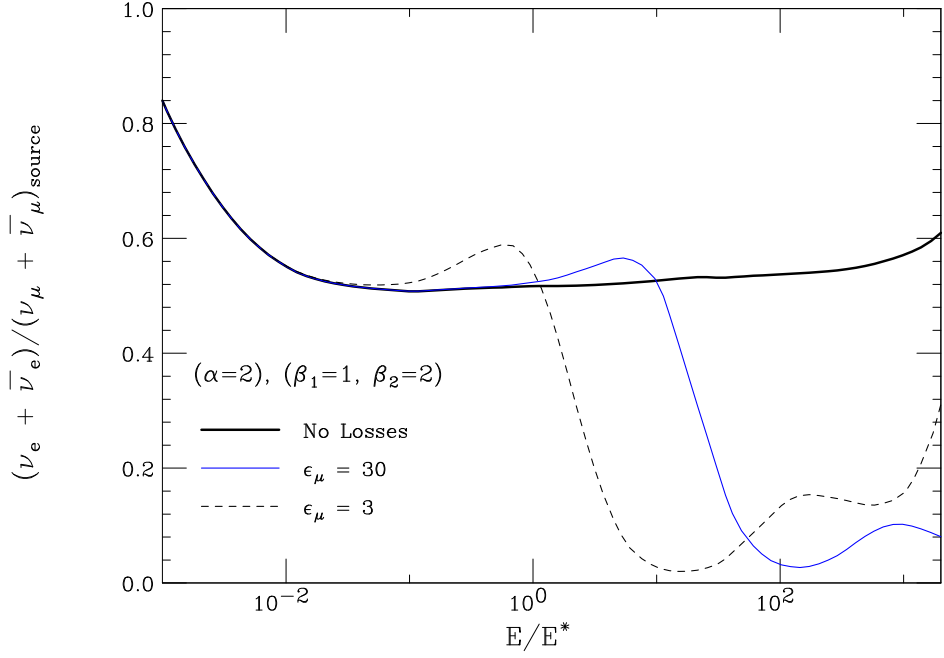,angle=90,width=12.5cm}}
\caption {
Ratio $(\nu_\mu + \overline{\nu}_\mu)/(\nu_e + \overline{\nu}_e)$  for 
neutrinos from a  GRB source. The  ratio   correspond
to the spectra shown in  fig.~\protect\ref{fig:nu1a}.
The three  curves are calculated   assuming
different  values of  the magnetic  field in the source
(and proton slope $\alpha = 2$, target photon
slopes $\beta_1 = 1$ and $\beta_2 = 2$).
The decrease  of the flavor  ratio  at high energy
for  large $B$ is  associated to the 
important synchrotron  losses of high energy muons.
\label{fig:nu2a}  }
\end{figure}

The flavor ratios  $R_{e\mu} = 
(\nu_e + \overline{\nu}_e)/(\nu_\mu + \overline{\nu}_\mu)$
that  correspond to  the   neutrino  spectra of 
the previous  figures
are  shown in figures~\ref{fig:nu2}  and~\ref{fig:nu2a}.
The  flavor  ratio is  energy  and model  dependent.
The $R_{e \mu}$  ratios  shown in figure~\ref{fig:nu2}  
correspond the  energy spectra
of fig.~\ref{fig:nu1}  and are 
 calculated  assuming that all energy
losses are negligible.
At the lowest  energy 
the ratio  $R_{e\mu}$  increases rapidly due
to the contribution of $\overline{\nu}_e$ from $n$ decay.
At higher  energy the ratio  $R_{e\mu}$  stays  close to the
value of 1/2, however  the expanded  scale
allows  to see that the ``naive'' result is not exact.
The value of $R_{e\mu}$,
for the  reasons   illustrated in 
section~\ref{sec:power},  are  correlated with 
the shape of the energy spectrum.
The  gradual softening of the neutrino energy distribution
with  growing $E_\nu$ 
is  reflected in a slow increase of $R_{e \mu}$.
Comparing the different  curves, one can also see that
the curve  calculated for $\alpha = 2.4$  
that gives the softest  neutrino  spectrum
correspond   (at sufficiently high energy) to the highest $R_{e\mu}$.

The effects  of   the existence of  significant
synchrotron losses are  very  important for  
the flavor ratio.
The main effect is the effective loss
of the   neutrinos from high energy muon
decay. This  results in a ``pile--up'' effects
that produced an {\em increase} in the ratio
for neutrino energies  below the 
synchrotron muon  energy  $\epsilon_\mu$. Above this  energy
the  ratio falls    steeply  reflecting the 
effective absence
of the main  source (muon decay) of electron neutrinos.
The flavor ratio  however does  not vanish  because
kaons  have decay modes   into  electron neutrinos,
and reaches a level of a few percent.
At very large energy    the role
of neutral  kaons   (that do not suffer  synchrotron losses)
become enhanced  and the  flavor ratio    grows
because  of the importance of the
$K^\circ_{e 3}$  decay  mode.

\begin{figure} [hbt]
\centerline{\psfig{figure=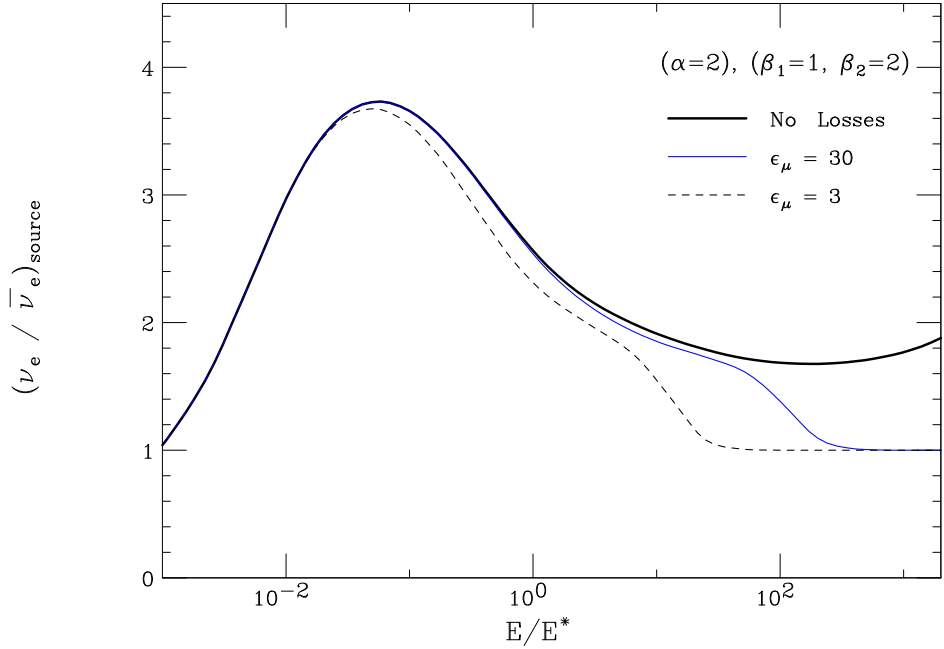,angle=90,width=12.5cm}}
\caption {
Ratio   $(\nu_e/ \overline{\nu}_e)$  at the source  for 
neutrinos from a  GRB source.
 The  ratio   correspond
to the spectra shown in  fig.~\protect\ref{fig:nu1a}.
At low  energy the ratio is low   because of the importance
of $\overline{\nu}_e$ from $n$ decay.  At higher  energy
the ratio  decreases   because of the increased
importance of $\pi^-$ production.
For  high magnetic  field, at  high energy
the dominant source of $\nu_e$ and  $\overline{\nu}_e$ 
becomes $K_L$ decay, and the   $(\nu_e/ \overline{\nu}_e)$
 ratio becomes unity. 
\label{fig:nu3}  }
\end{figure}

\begin{figure} [hbt]
\centerline{\psfig{figure=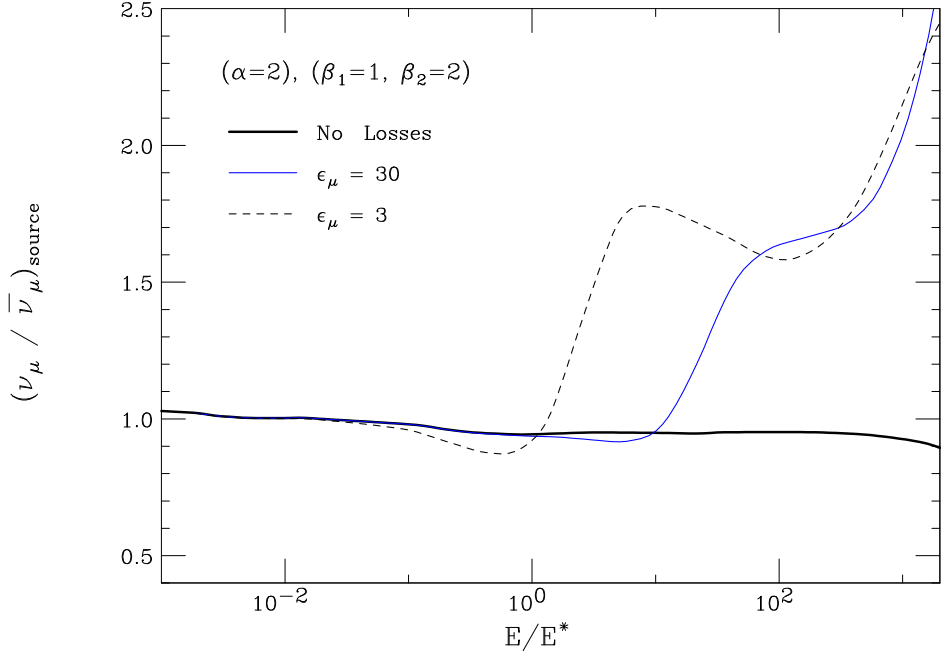,angle=90,width=12.5cm}}
\caption {
Ratio $(\nu_\mu/ \overline{\nu}_\mu)$   at the source 
for neutrinos from a GRB source.
The  ratio   correspond
to the spectra shown in  fig.~\protect\ref{fig:nu1a}.
The  three  curves correspond to three
different values of the magnetic  field in  the source.
At low  energy
(and at all energies for  small $B$)
 the ratio is unity, reflecting 
the fact that  the chain decay of a charged pion produces
a $\nu_\mu$ and a $\overline{\nu}_\mu$.  
In the presence of a large magnetic  field,   the neutrinos
from  muon decay are suppressed, and the flavor ratio   is  determined
by the $\pi^+/\pi^-$  ratio. At the highest energies
charged  kaon decay becomes the dominant 
source of $\nu_\mu$ and $\overline{\nu}_\mu$ and the  flavor ratio  reflects
the $K^+/K^-$ ratio.
\label{fig:nu4}  }
\end{figure}

Figure~\ref{fig:nu3}  and
  figure~\ref{fig:nu4}  show
the neutrino/anti--neutrino ratios
$\nu_e/\overline{\nu}_e$ and $\nu_\mu/\overline{\nu}_\mu$
calculated   for the same
three situations
present in   fig.~\ref{fig:nu1a} and~\ref{fig:nu2a}
(that is  slopes $\{\alpha, \beta_1, \beta_2\} = \{2,1,2\}$
and    three  values of the magnetic field
that  corresponds  to $\epsilon_\mu =  \infty$, 30 and~3).

The $\nu/\overline{\nu}$ ratios   are difficult
to  measure  because     the detectors
have  no capability to measure the charge 
of the   final state  charged  lepton in 
charged current neutrino interactions.
The most attractive idea  
\cite{Anchordoqui:2004eb}
to estimate
the    $\nu_e/\overline{\nu}_e$  ratio  is the  comparison
of the event rate   at and near the 
``Glashow  resonance''  \cite{glashow} at 
$E_\nu \sim  M_W^2/(2 \,m_e)$. The resonance
is  present  only  for 
 $\overline{\nu}_e$    in   reactions  such as
$\overline{\nu}_e + e^- \to  {\rm hadrons}$ 
that  can proceed   via   the   formation of a 
$W^-$  boson  in the $s$ channel, 
while non--resonant events proceed
via the  normal charged  current interaction  on nucleons.

The   study of $\nu_e/\overline{\nu}_e$ ratio  is  particularly
interesting because it   has  been 
proposed  
\cite{Anchordoqui:2004eb} as a method  to  distinguish 
$pp$ and $p\gamma$ interactions as the  source
of the neutrinos.
The  idea   is that  
the  $\nu_e/\overline{\nu}_e$ ratio  at the source   is
close to unity for a $pp$ source,
when the interactions  produce  approximately equal  numbers
of  positive and  negative pions (that decay into  $\nu_e$ and
$\overline{\nu}_e$),    and   is  very large
(or  actually diverges)   
for $p\gamma$ interactions,  when  most  interactions
happen close to threshold, and the  cross section
is  dominated  by single  pion production 
(via  the $\Delta$  resonance), and therefore   only $\pi^+$ production
is  present.
This argument is  qualitatively   right,
  but quantitatively incorrect, in fact as   discussed  
above, in general   increasing $E_p$
a   broader  range of  c.m. energies becomes 
possible,  and this results in a  growing  contribution of 
negative pions  to neutrino  production. In certain
circumstances  also   the contribution of kaons  can become
important.
This is  illustrated in fig.~\ref{fig:nu3} 
that  shows the  $\nu_e/\overline{\nu}_e$ ratio.
At the lowest energies the flux   is dominated
by the $\overline{\nu}_e$ from $n$ decay, and the $\nu_e/\overline{\nu}_e$ ratio 
is therefore small.
When  the contribution of neutrinos  from 
 pion decay becomes dominant the 
ratio  $\nu_e/\overline{\nu}_e$ is  significantly larger than one,
because  positive pions
are more abundantly  produced  than  negative ones.
In fact  below  the threshold for two pion 
production 
the ratio $\pi^+/\pi^-$ diverges.
With  growing energy  the ratio $\pi^+/\pi^-$  decreases 
because of the  growing  importance of  multiple
pion  production.
This is reflected  in a decrease  of the 
$\nu_e/\overline{\nu}_e$ ratio  that at large energies  becomes  $\sim 2$.
If  a  large magnetic  field
is present, 
at large  energy the  synchrotron   energy losses  for muons
become important and neutrinos  from muon decay
are suppressed.
The decay of neutral  kaons ($K_L$) becomes
the dominant  source of $\nu_e$ and $\overline{\nu}_e$, 
and the ratio  $\nu_e/\overline{\nu}_e$  becomes unity.

The ratio $\nu_\mu/\overline{\nu}_\mu$ (shown in fig.~\ref{fig:nu4})
is approximately unity  at low  energy,  reflecting the fact
that each charged  pion, after chain decay,
contributes a $\nu_\mu$ and a $\overline{\nu}_\mu$.
The existence of important synchrotron losses
can modify this  quite robust  prediction.
At high energy, when  muons energy losses
become significant, only the
neutrinos from direct  pion decay are  important, and
the ratio $\nu_\mu/\overline{\nu}_\mu$  is   approximately equal
to the $\pi^+/\pi^-$. 
At larger energy,   charged  kaons   become
 the dominant  $\nu_\mu$ and $\overline{\nu}_\mu$  source
and the   $\nu_\mu/\overline{\nu}_\mu$ ratio 
increases reflecting the  higher $K^+/K^-$ ratio.

\begin{figure} [hbt]
\centerline{\psfig{figure=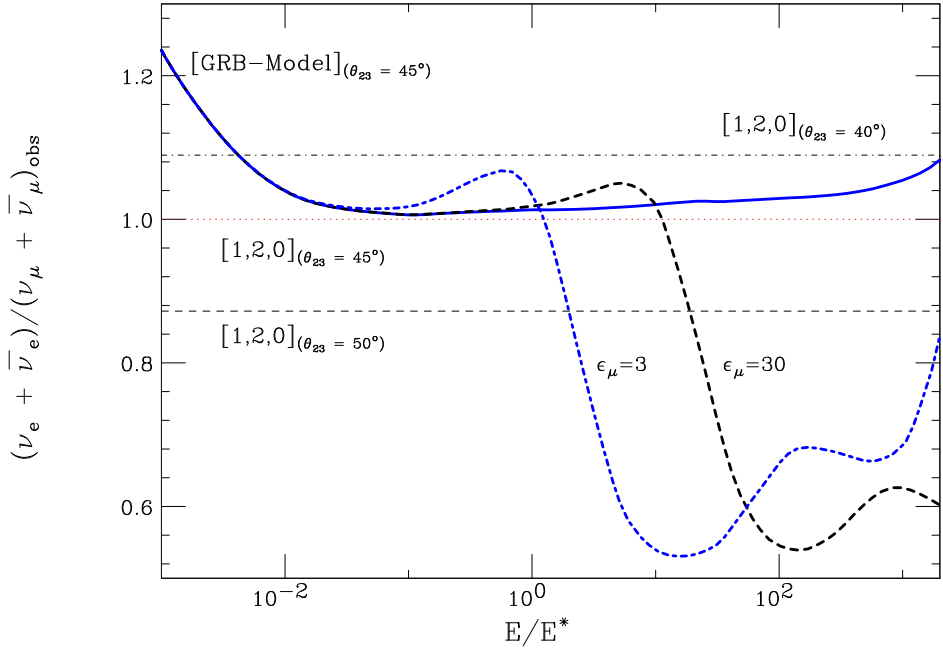,angle=90,width=12.5cm}}
\caption {
Observable ratio $(\nu_\mu + \overline{\nu}_\mu)/(\nu_e + \overline{\nu}_e)$ 
after  propagation 
for the  neutrinos from a GRB source.
The  solid lines  corresponds to the 
source models  shown in fig.~\protect\ref{fig:nu1a},
and~\protect\ref{fig:nu2a}. The three curves  differ for the
value of the  magnetic field   that corresponds to $\epsilon_\mu = 3$,
30 and $\infty$  (no field). 
The neutrino oscillation parameters  have been chosen
as $\theta_{23} = 45^\circ$,
$\theta_{12} = 34^\circ$
and  $\theta_{13} = 0$.
The thin dotted are calculate
for the ``naive'' assumption of a  source
composition  $[\nu_e,\nu_\mu, \nu_\tau] = [1,2,0]$.
The dotted  line   assumes the same  mixing parameters
as  above. The Dashes (dot--dashed) line
has $\theta_{23} = 50^\circ$  ( $40^\circ$).  
\label{fig:obs}  }
\end{figure}

In fig.~\ref{fig:obs}  we  show the 
$(\nu_e/\nu_\mu)$  ratio  for  an Earth observer.
To propagate  neutrino we have used standard oscillations
with mixing  parameters  $\theta_{23} = 45^\circ$,
 $\theta_{12} = 34^\circ$
and $\theta_{13} = 0$.
In the figure we have also shown
the same  observable   ratio
for   a  source
with  the ``naive'' 
composition  $[\nu_e,\nu_\mu, \nu_\tau] = [1,2,0]$.
For the ``best fit''  values  of the mixing
parameters   the  observable $(\nu_e/\nu_\mu)$   
has  the well known  value of unity.
Changing the value of the $\theta_{23}$ angle
by $\pm 5^\circ$ the   ratio  changes by 
$-13\%$ ($+7\%$).
Similarly   for   $\theta_{13} = 5^\circ$
the ratio  take   values in the interval (0.95,1.04)
depending on the value of the phase $\delta$.
This  figure  illustrates the  interplay between 
astrophysical and  particle physics  uncertainties
(also   shown before in equations
(\ref{eq:rflav1}), 
(\ref{eq:rflav2}) and   (\ref{eq:rflav3})).
In order  to  perform a measurement of $\theta_{23}$
one  clearly needs an  accurate control of the initial  
flux composition.

The description of the  emission of neutrinos
from GRB that we have used in this  section 
following reference  \cite{Waxman-grb} is 
very simplified,  and  could be improved
in many aspects.  The results  of our
calculation  show  that  one  obtains  non  trivial
and  intriguing  energy dependences  of the  flavor ratios,
that could reveal important informations about the 
source.  
This qualitative observation  is   
of a much  more general validity
and one  expects
that in  most cases, and   also using 
more realistic  descriptions  of  the sources,
one should  find  non  trivial structures in 
the energy spectra and flavor composition
of the astrophysical neutrino signals.

\section{Experimental Flavor  identification}
In this   work we will  not  consider 
in detail   the 
crucial problem  of  the experimental  determination
of the flavor ratios.
Several  experimental  approaches 
for the detection are at present  being
developed.  The concept  that is  in the  most advanced
stage is  a large volume (cubic Kilometer) 
ice or water Cherenkov  detector, with  photon detectors
distributed inside  the volume, as 
originally proposed   by  the Dumand  group 
\cite{Dumand}  and now  in  different  stages of development
 at the South Pole
\cite{Amanda,IceCube},  in the Baikal lake \cite{Baikal}
and in the mediterranean sea \cite{Nestor,Antares,Nemo}.
Several other  alternative techiques are also  being developed
\cite{Learned:2003}, that includes radio \cite{Rice,Anita}, 
acoustic
and air shower  \cite{Letessier-Selvon,euso-nu}  detection methods. 

A  very imporant problem is of course
the amount of data  that is required 
in order  to reduce the statistical
errors to  the level needed for a meaningful measurement.
Today we  only  have  upper limits  for the astrophysical 
neutrino fluxes, optimistically one can expect that 
these fluxes will be soon  discovered just below the
current upper limits,  however even in this case
the  event rates  will remain small, and  one will  need
very long data  taking  periods,
or  much larger mass detectors  to collect sufficient data
\cite{vissani,meloni}.

In this section we want to  stress the point
that  in all the experimental methods  that are
envisaged, the measurement  of the flavor ratio
is  connected with a  determination of
(or a theoretical assumption on)
the shape of the  energy distribution
of the neutrino fluxes.
This  happens  because   the   detection
methods of different neutrino  flavors
have efficiencies with   different 
dependences on $E_\nu$.
If the statistical errors are   sufficiently small
(with  very  large mass  detectors), and 
having achieved perfect control of  the 
 detector  acceptances and efficiencies, 
the uncertainties in the determination of the
neutrino spectral  shapes
will  remain as  the dominant source
of  error in  the measurement of the flavor ratios.

To illustrate  the interplay
between the spectral  shapes and the determination
of the  flavor ratio one can  consider
the case of
the Km$^3$  type  Cherenkov telescopes.
In these detectors one expects 
to find  different classes of
neutrino induced events \cite{Beacom-flavor}.  
\begin{itemize}
\item [(i)] In ``Track Events'' a single 
up--going  muon is seen entering the detector.
The energy of  the muons 
can be estimated    from the amount of  Cherenkov light
produced  by  the photons and $e^\mp$ pairs 
radiated  from the muon  track.
\item [(ii)] In ``Shower Events''
one observes   the release of  a large  amount
of  Cherenkov light 
generated by   the shower induced by a  neutrino interaction
in the detector volume. 
The amount of light can be   translated in
the   estimate of a  visible energy
release $E_{\rm vis}$.
\item [(iii)] In ``Double Bang Events''   
\cite{double-bang,Fargion:1997eg}
one  detects  two  distinct energy releases 
inside  the detector  volume, with space
and time  separations  that are
consistent  with the  propagation
and decay of  an energetic  tau  lepton.
\end{itemize}
Events in class (i)  are associated  with 
the charged current  (CC) interactions  of
muon neutrinos  below  the detector.  Events in  class (ii)
can  be produced  by several  types  of neutrino
interactions. Electron  (anti)neutrino  CC  interactions
are  the main source of   
shower events, 
there are also   contributions
from the  neutral current interactions of
all neutrino  types, and in general also
from the  CC  interactions of
$\nu_\mu$ and $\nu_\tau$.
Finally  events in class (iii) 
are  generated by $\nu_\tau$  CC interactions.
Clearly, the ratio  of the frequencies of 
these three classes of  events  does  give
information about the  flavor  composition of
the neutrino  fluxes.
In fact the ratio of the event rates for
``Shower''  and ``Tracks''  events  
has been proposed as the best method 
for a measurement of the $\nu_e/\nu_\mu$    flavor  ratio.

The frequency of ``Track'' events  
can be  studied as a function of the visible
 muon energy $E_\mu$.
The differential event rate  can be calculated as:
\begin{equation}
\frac{dN_{\rm Tracks}}{dE_\mu\; d\Omega_\oplus \; dt} (E_\mu, \Omega_\oplus) =
A_{\rm det}^{\rm Tracks} (E_\mu, \Omega_\oplus) ~ 
\sum_{\nu_{\mu}, \overline{\nu}_{\mu}} 
~ \int dE_\nu ~
 \phi_{\nu} (E_\nu, \Omega_\oplus) ~ \frac{dY_{\nu \to \mu}}{dE_\mu} (E_\mu;~E_\nu) 
\label{eq:tracks}
\end{equation}
where $A_{\rm det}^{\rm Tracks} (E_\mu, \Omega_\oplus)$ is
the detector  effective area for the
detection of  muons  with energy $E_\mu$  from
the direction $\Omega_\oplus$, 
$dY_{\nu \to \mu}/dE_\mu (E_\mu; ~E_\nu)$ 
is the 
``muon yield'', that is the  probability that
a  $\nu_\mu$ ($\overline{\nu}_\mu$) of  energy $E_\nu$   
produces a  $\mu^\mp$
of energy $E_\mu$  at the detector.
In this  equation   we  have  assumed
that the muon and the neutrino are collinear.
Even for an individual  source, the   energy spectrum  
will depend on the zenith angle $\theta_\oplus$,  
because of the effects of neutrino absorption in the Earth.  
This  procedure  hower  implies  a  good understanding of the
neutrino energy spectrum.

The rate of  ``Shower''  events  with 
a visible energy  release   $E_{\rm vis}$
can be estimated as the   convolution
\begin{eqnarray}
\frac{dN_{\rm Showers}}{dE_{\rm vis} \; d\Omega_\oplus \; dt}
 (E_{\rm vis}, \Omega_\oplus)
& \simeq  &  \frac {M_{\rm det} (E_{\rm vis})}{m_p}
~ \Big \{
 \sum_{\nu_{e}, \overline{\nu}_{e}} 
\left [ \phi_{\nu} (E_\nu, \Omega_\oplus) ~ \sigma_\nu^{\rm cc} (E_\nu)
\right ]_{E_\nu \simeq E_{\rm vis}} 
\nonumber \\[0.2 cm]
 & + & 
 \sum_{{\rm All}~\nu} 
 \int dE_\nu ~ \phi_{\nu} (E_\nu, \Omega_\oplus) \int_0^1 dy~
\frac{d\sigma_\nu^{\rm nc}}{dy} (y,E_\nu) \;
\delta [E_{\rm vis} - E_\nu \, y]~~~~~
\label{eq:showers}
\\
~ & ~ & \nonumber \\
~ & + & 
\left [ (\nu_\mu, ~\overline{\nu}_\mu) ~{\rm CC} 
\right ]
+ \left [ (\nu_\tau, ~\overline{\nu}_\tau) ~{\rm CC} 
\right ]
~~\Big \}~.
\nonumber 
\end{eqnarray}
In this equation 
$M_{\rm det} (E_{\rm vis})$ is the  effective detector
mass;
the first line is the contribution
from the charged  current interactions of electron
(anti)--neutrinos, in  this case 
the visible energy,  in  first approximation
(neglecting differences  in the  light to energy
conversion for e.m. and  hadronic showers)
equals the   initial  $\nu$ energy;
the second line 
is the contribution from  neutral current  interactions,
in this  case the 
visible  energy is   only a fraction of the initial  energy
because the  neutrino in the final state
carries  away   an invisible  energy
$E_{\nu} \, (1-y)$.
The contributions  of $\nu_\mu$  and $\nu_\tau$ CC
interactions  to the shower event  rate  in
(\ref{eq:showers}) has   been  left   implicit.
A  fraction of these  interactions can   probably be identified
and  eliminated
selecting  events with  a muon in the  final state, or
a double  bang structure, but   the rest will
contribute to the ``shower''  class of events.

From (\ref{eq:tracks}) 
and (\ref{eq:showers})  
it is  clear that the  
energy response  for the two classes of events
is different, with the ``Track'' events  selecting
larger  $E_\nu$.  Therefore the 
ratio for the integral rates
for ``Shower'' and  ``Tracks''    events 
is  strongly dependent on the spectral shape of the fluxes.
Use of the differential  distributions in
$E_\mu$ and $E_{\rm vis}$ allows in principle
to unfold the  shape of  the neutrino energy spectra,
however an  important  difficulty is that 
a  narrow  interval  in the muon energy
$E_\mu$ (that can  however be  measured  only with 
modest  resolution) corresponds to a very broad  
interval in $E_\nu$. Also
$E_{\rm vis}$ in  shower events is not in a one to one
correspondence with $E_\nu$.
The  unfolding procedure 
has therefore  some  significant limitations.

As an illustration of these problems, 
one can  look back at the original discovery
of neutrino oscillations \cite{Kamiokande,IMB,SK} 
with atmospheric neutrinos.
The  evidence
collected by the SK  \cite{SK} experiment
for subGeV  fully contained events ($E_{\rm vis} \le 1.33$~GeV)
implied   comparing the  ratio
of the   observed rates of 
$e$--like and $\mu$--like events   with a MC prediction 
$(\mu/e)_{\rm MC} \simeq 1.50$. This prediction  is significantly 
smaller than  the no--oscillation ratio 
$(\nu_\mu+ \overline{\nu}_\mu)/(\nu_e + \overline{\nu}_e) \simeq 2$ for
the combination of several reasons, related to the  difference
in detector  acceptances  for $e$--like and $\mu$--like  events.
The 
experimental cuts   select for the $e$--like events
a broader energy range,
both at the low and high energy ends
(at low  energy  because electrons have  a larger  yield of
Cherenkov photons, and at high energy   because electrons have a 
shorter range and are more easily contained).
For  this reason the MC
prediction does  depend  on  the  shape
of the neutrino fluxes. 
For these sub--GeV events,   
the    intervals of  $E_\nu$  
for the two classes of $e$--like and $\mu$--like events
are very  similar, and therefore the 
associated systematic uncertainty
in the $(\mu/e)$ prediction 
is relatively small, but it
remains as  a  relevant  source of error.
 
The comparison of the   rates  of  classes of events  
that correspond to   very different  atmospheric  neutrino energy,
such as samples of fully contained  $e$--like events  
and  upgoing muons
(that are produced  by $\nu_\mu$ charged current interaction
with a median energy of order 100~GeV) 
also  allows in principle to verify 
the existence of $\nu_\mu \leftrightarrow \nu_\tau$   transitions.
However in  this case  the systematic  uncertainty in  the prediction 
of the shape of the energy distribution   is  much larger,
and in practice this  method  gives only marginal  evidence for
the existence of a suppression of the  upgoing   muon rate.

In conclusion, the extraction of the
$e/\mu$  flavor  ratio  from
the  comparison of the rates of ``Track'' and ``Shower''  events
in high energy neutrino telescopes 
requires    excellent  control of the detector
performances, and    a  sufficiently precise knowledge  of the
shapes of the  neutrino  spectra.
The determination of these shapes  could be  obtained
from the data  themselves (the ``unfolding method'')
or with the help of  reliable theoretical models  for the source.
In both cases  the task appears as remarkably  difficult.

\section{Conclusions}
The study of astrophysical  neutrinos
can be used for  two distinct goals: the  understanding
of their  sources  and the investigations 
of  the neutrino properties.
These two goals  are in  a sense contradictory.
The investigation of the  neutrino fundamental properties 
requires a comparison of  the observations  
with  expectations that must necessarily be  based
on a  sufficiently accurate  understanding of the source.
On the other  hand the  flavor composition
of  a neutrino signal  encodes information
about the source  properties,   
but in order to
extract this information one needs to know the properties
of  flavor transitions   over  galactic or cosmological
distances with  sufficient  accuracy.

Several recent works
\cite{Pakvasa:2004hu,Beacom-flavor,Barenboim:2003jm}
have  speculated   that it is possible
to discovery new  neutrino  properties,
or even  measure precisely   the parameters of
the already established  standard oscillations
\cite{astro-standard}
from the observation of  future astrophysical
neutrino sources 
 assuming  that the flavor  ratios
at the source  are robustly predictable.
The  often  repeated  argument is that
a ``standard neutrino  source'' is  
dominated by charged pion decay, and  
therefore    emits the  different   flavors 
(summing over $\nu$ and $\overline{\nu}$)  with relative intensities:
[$\nu_e, \nu_\mu, \nu_\tau] \simeq [1, 2, 0]$.
This   statement  however 
must be understood as  a first approximation
of  only  limited   validity. 
First of all  it   requires  that the energy losses  of
the  muons   created in charged pion decay are  negligible.
This  crucial assumption   is  satisfied
for  several   but not all  of the proposed  neutrino sources.
Even for  ``thin environment  sources''
(where    the density 
of   ordinary matter  and  magnetic 
field  is  sufficiently  low  so that
the  energy  losses
for   all secondary unstable particles can be safely neglected)
the   flavor  ratios are    (slowly varying) 
functions of the   neutrino  energy, with values
that are  determined by  the shape of  the neutrino spectra
and by the nature of target material.
This  dependence arises  from two  effects:
the first is that  in general   the contributions
of other  weakly  decaying particles 
(kaons  and  in some  cases also neutrons)  is  not
negligible,
the  second  is that, even if charged pions are
the only significant neutrino  source,
the three neutrinos  that  are emitted in a $\pi^\pm$   chain decay
have different  energy spectra. 
Folding these  decay spectra 
with the energy distribution of the parent particles, one obtains
slightly different  shapes, and  therefore
energy  dependent flavor  ratios.
In the special case where the parent pion spectrum  is a  power law,
the neutrino energy distributions  have  also a  power law form
with the same slope, and the $\nu$ flavor ratios are constants.
As a  (phenomenologically important)  example, 
 the neutrino  fluxes  with  spectrum
 $ \propto E_\nu^{-2}$  emerging from a  ``thin'', 
charged pion source,  
have  relative intensity  (at the source)
[$\nu_e, \nu_\mu, \nu_\tau] \simeq [1, 1.86, 0]$, a 7\%  deviation
from the naive  expectation.

In general, 
the  combined effects  of  different  shapes
 for  the $\nu$  energy spectra,
and  different  contributions of  the relevant weakly decaying particles
correspond to  variations of 
order 10\%   of the $\nu_e/\nu_\mu$  ratio
(in some  special  circumstances even larger).
The size of these corrections  is    however
comparable to 
the effects on the  observable  flavor  ratios 
due to the present uncertainties  
in the values  of the standard  oscillation parameters
(for a  quantitative  illustration of this  point
see equations
(\ref{eq:rflav1}), 
(\ref{eq:rflav2}) and   (\ref{eq:rflav3})).
The  very  ambitious   program  
\cite{astro-standard}  to attempt a precision 
measurement of $\theta_{23}$ or a determination of $\theta_{13}$
with astrophysical  neutrinos  cannot  therefore  ignore 
these problems, and   requires 
a very  precise  control of the structure and
properties of the neutrino sources.

Much larger effects on the neutrino  flavor ratio 
arise  from situations  where the  environment of
the  source contains  large  energy densities
in magnetic  field, or in ordinary matter.
In the  presence of sufficiently  large magnetic  fields,
such as  those that  could exist in a GRB fireball,
high energy muons  can lose most of their energy
to  synchrotron radiation \cite{Rachen:1998fd,Kashti}
 and their  contribution
to the neutrino  fluxes becomes strongly suppressed.
This  results in a   very small $\nu_e/\nu_\mu$ ratio,
with a value determined by the size of  the kaon contribution
(that  generate  electron neutrinos in direct decay modes).
At higher energy also pions  suffer
important   synchrotron losses, and kaon decay
become the   dominant  $\nu$ source   for all flavors. 
In some  extreme  circumstances  neutral kaon decay
will remain as the  only  significant neutrino source
with a $\nu_e/\nu_\mu$ ratio  at the source of order unity. 

In some  circumstances  it is possible
that interactions with ordinary matter 
play a major r\^ole.
If the density is sufficiently large 
high energy pions  reinteract  before decaying.
This  supresses the neutrino fluxes and  enhances  the contribution
of kaon  decay. Summing over charged and neutral kaons,
this  also results in a significantly  reduced  $\nu_e/\nu_\mu$ ratio.

The prediction of the neutrino flavor ratios
in these circumstances  (strong magnetic field
or large densities of ordinary matter) is going to be
very difficult, because  the assumption of
a ``single volume''   homogeneous source is very likely
a poor (and in any case not  demonstrable) assumption.
The neutrinos 
will likely emerge with different   spectra and   flavor ratios
from different  regions of the source, and for example
fitting  the data  including a value of the field
$B$, and/or  a matter density $n_{\rm gas}$ is  likely to be
only a poor approximation.

The existence of  neutrino  sources  dominated
by neutron decay are  also a   theoretical possibility, in  situations
where   photodisintegration of relativistic nuclei
interacting on a  radiation field 
is possible 
while   pion  photoproduction is  kinematically
forbidden. 
If such sources  exist, it will
however be very difficult   to use these
fluxes for   studies of the neutrino flavor transitions
because   in general one expect them
to have an important ``contamination''  of muon 
neutrinos in their  initial  flavor composition.
In fact  it is unnatural to have
the pion photoproduction   to be absolutely forbidden,
because 
the  energy thresholds for nuclear photodisintegration
and pion production  differ only by  one order of magnitude.  
The size  of the contribution
of pion decay to the neutrino fluxes in these circumstances
is  difficult to control.

Even considering   all the caveats of  our discussion,
there are some  circumstances where 
observations  of astrophysical  neutrinos  could
lead  to the  discovery  that 
new physics  is involved  in   the propagation of neutrinos
over astrophysical  distances.
In particular the investigation of the existence
of neutrino  decay \cite{nu-decay}
is a  very attractive possibility.
The observation of the absence 
(or very strong  suppression) of
a  $\nu_e$  component  in the measured fluxes
cannot be explained  with  even  extreme  variations
of  the source  astrophysical properties,
and is  on  the  contrary  predicted 
(for  any  type of source) in   
a  neutrino decay model  
with inverse mass hierarchy.
For  the   normal  mass hierarchy
the neutrino decay model  predicts 
a large $(\nu_e/\nu_\mu)_{\rm obs}$  ratio
between 3 and 6.  
If  the neutrinos  have standard properties,
ratios  $(\nu_e/\nu_\mu)_{\rm obs}$    much larger than unity
are only possible  (and  limited to 
$\lesssim 4$)  for sources  dominated by $\nu_e$   emission.
Therefore   from  the observation   of a large
$(\nu_e/\nu_\mu)_{\rm obs}$ ratio  it could be possible
to   infer the   existence of   $\nu$  decay 
in the normal mass  hierarchy case.
This   could require  some  understanding
of the  properties of the $\nu$ source.

In the general  case the observation of
flavor ratios  very  different from the 
``standard  source'' values (relative  intensities 
$[\nu_e, \nu_\mu, \nu_\tau]_{\rm obs} \sim [1,1,1]$) 
 could  be  interpreted as  evidence
for some  form of  new physics  in neutrino  propagation,
but also, and probably more economically, as  an  indication
of   the existence of some unexpected  properties in 
the $\nu$ source.
Resolving these ambiguities   is in principle
possible, but certainly not easy, and requires
a careful interpretation  of the multi--messenger
(photons, neutrinos and possibly  cosmic rays) and
multi--wavelength observations.

\newpage

\clearpage

\begin{figure} [hbt]
\centerline{\psfig{figure=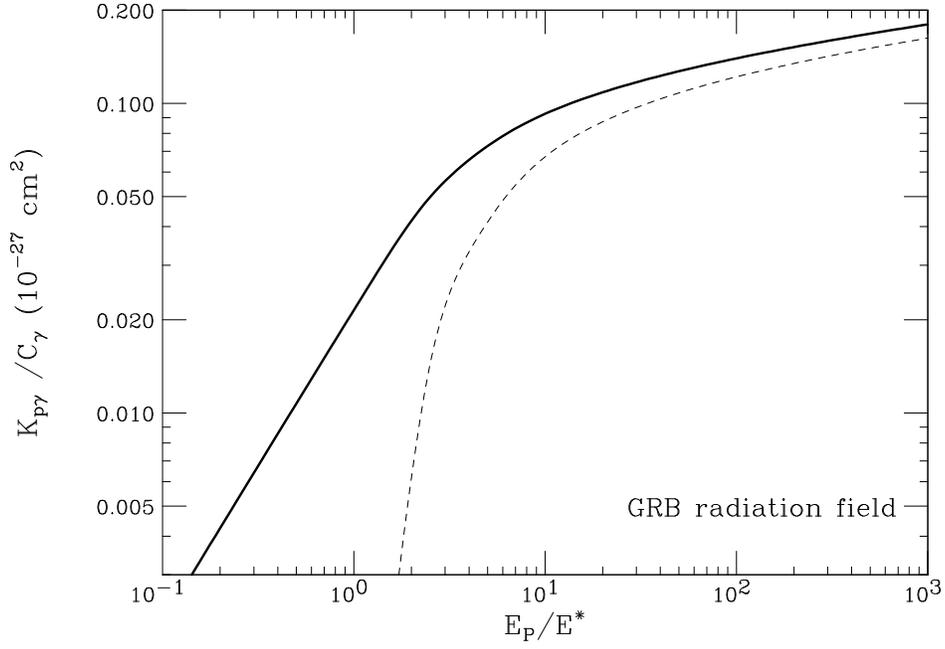,angle=90,width=12.5cm}}
\caption {
Proton interaction rate  in  a  radiation field
of  the form (\protect\ref{eq:grb-spectrum}).
The energy $E^*$ 
(defined in equation (\ref{eq:estar}))
corresponds to the threshold  proton energy 
for  interaction with  photons   with energy
$\varepsilon_{\rm b}$.  The dashed line 
neglects the interactions with all photons
with energy  $\varepsilon > \varepsilon_{\rm b}$.
\label{fig:k_pgamma}  }
\end{figure}

\begin{thebibliography}{200}
\bibitem{photo-mc} The details  of the Montecarlo code
will be presented elsewhere.
\end{thebibliography}
\end{document}